\documentclass[useAMS,usenatbib]{mn2e}
\usepackage{graphicx}
\usepackage{natbib}

\newcommand{\source}{LS~5039}

\newcommand{\psr}{PSR~B1259-63/SS2883}
\newcommand{\hess}{HESS}

\title[On the formation of TeV radiation in \source]{On the formation of TeV radiation in \source}

\author[Khangulyan, Aharonian \& Bosch-Ramon]
       {D.Khangulyan\thanks{E-mail:dmitry.khangulyan@mpi-hd.mpg.de}${}^1$, F.Aharonian${}^{2,1}$ \& V.Bosch-Ramon${}^1$\\
        ${}^1$Max Planck Institut f\"ur Kernphysik, Heidelberg, Germany\\
	${}^2$Dublin Institute for Advanced Studies, Dublin, Ireland}

\begin{document}

\date{Accepted. Received; in original form}

\pagerange{\pageref{firstpage}--\pageref{lastpage}} \pubyear{}

\maketitle

\label{firstpage}

\begin{abstract}
The recent detections of TeV gamma-rays from compact binary systems
show
that relativistic outflows (jets or winds) are sites 
of effective acceleration of particles up to multi-TeV energies. 
In this paper, we discuss the conditions of acceleration and radiation of ultra-relativistic electrons
in \source, 
{ the gamma-ray emitting binary system for which the highest quality TeV data are available.}
Assuming that the gamma-ray emitter is a jet-like structure, we performed 
detailed numerical calculations of the energy spectrum 
and lightcurves accounting for the acceleration efficiency, 
the location of the accelerator,
the speed of the emitting flow, the inclination angle of the system, 
as well as specific features related to anisotropic 
inverse Compton (IC) scattering and pair production. 
We conclude that the accelerator 
should not be deep inside the binary system unless we assume a very efficient  acceleration rate.
We show that within the IC scenario both the gamma-ray spectrum and 
flux are strongly orbital phase dependent. Formally, our model 
can reproduce, for specific sets of parameter values, the  
energy spectrum of gamma-rays reported by \hess{} for 
wide orbital phase intervals. 
However, the physical
properties of the source can be constrained only by observations 
capable of providing detailed energy spectra for narrow orbital phase intervals
($\Delta\phi\ll 0.1$). 
\end{abstract}


\section{Introduction}

\source{} is a binary system consisting of a very bright star and a compact object (a neutron star/pulsar or
a black-hole). The source emits X-rays (see e.g. { \cite{bosch07} and references therein;} 
\cite{goldoni06}; \cite{derosa06}), and
presumably also MeV \citep{strong01} and GeV \citep{paredes00,paredes02} gamma-rays. Recently, \source{} has been
detected in very high energy (VHE) gamma-rays by the \hess{} array of atmospheric Cherenkov telescopes
\citep{hess_ls5039_05}. The TeV radiation of the source is clearly modulated with a period $3.9078\pm 
0.0015$~days \citep{hess_ls5039_06}, which perfectly coincides with the orbital period of the object \citep{casares05}. 

The nature of the compact object in \source{} is not yet firmly established, given the uncertainty related to the orbital inclination angle,
$13^\circ<i<64^\circ$, which does not allow precise estimate of its mass. { The optical line analysis, 
together with the assumption of orbital pseudosynchronization favour a
rather small inclination angle, $ i\simeq25^\circ$ \citep{casares05}}. 
This would be an indication of a rather high mass for the compact object, $M\simeq 3.7\
M_\odot$, and therefore could be interpreted as an evidence for its black-hole nature. In such a
case, one would expect the realisation of the {\it microquasar} scenario, in which the nonthermal processes take place in a jet related to an
accreting compact object.  The original classification of \source{} as a microquasar was based on its extended radio emission features
\citep{paredes00}. In the framework of this scenario, a number of models have been proposed to explain the TeV gamma-ray emission of
\source{} based on both leptonic and hadronic interactions (e.g. \cite{dermer06,aharonian06,paredes06,bednarek07}). At the same time, a
scenario in which the nonthermal radiation is related to an ultrarelativistic pulsar wind, remains an alternative option \citep{dubus06}. In
this scenario, \source{} would { behave in a similar manner to the binary pulsar systems} (e.g. \psr), where the production site of gamma-rays is (most
likely) related to the pulsar wind termination shock \citep{maraschi81,tavani97,kirk99,khangulyan_psr}. Whereas in the microquasar scenario, the particle
acceleration and gamma-ray production are possible throughout the entire jet (i.e. both inside and outside of the binary system), { in} the
standard pulsar wind model the particle acceleration and radiation take place well inside the system \citep{dubus06}, namely at distances 
from the compact object much smaller 
than the separation between the stars, i.e. $R_{\rm orb}=(1.4-2.9)\cdot10^{12}{\rm cm}$ \citep{casares05}.

The discovery of modulated TeV gamma-ray emission in \source{} is a strong indication that gamma-rays are produced { close to} the binary system. On
the other hand, if the gamma-ray production region is located inside or very close to the system, { we expect} distinct photon-photon absorption
features caused by the interaction between the gamma-rays and the stellar radiation field. { Within the leptonic models for gamma-ray production}, two
other effects have { a} strong, direct or indirect, impact on the formation of the gamma-ray spectrum. One effect is related to the anisotropic IC
scattering; the other effect is related to the maximum energy of relativistic electrons determined by the balance between the acceleration and
radiative cooling rates. Finally, the production and gamma-ray absorption processes should be coupled with an appropriate treatment of the particle
cooling and propagation (e.g. particle diffusion and advection along the jet). Thus, in the context of leptonic models, the spectral shape of the
gamma-ray emission, in particular its orbital phase-dependence, contains unique information about the location of the particle acceleration and
gamma-ray production regions. 

In what follows, we show that the spectral and temporal features of the VHE gamma-radiation reported by \hess{} from \source{} can be
reproduced with a leptonic model that invokes anisotropic IC scattering of electrons and absorption of gamma-rays under certain
requirements concerning the location and efficiency of the electron accelerator.

\begin{figure}
\includegraphics[width=0.5\textwidth]{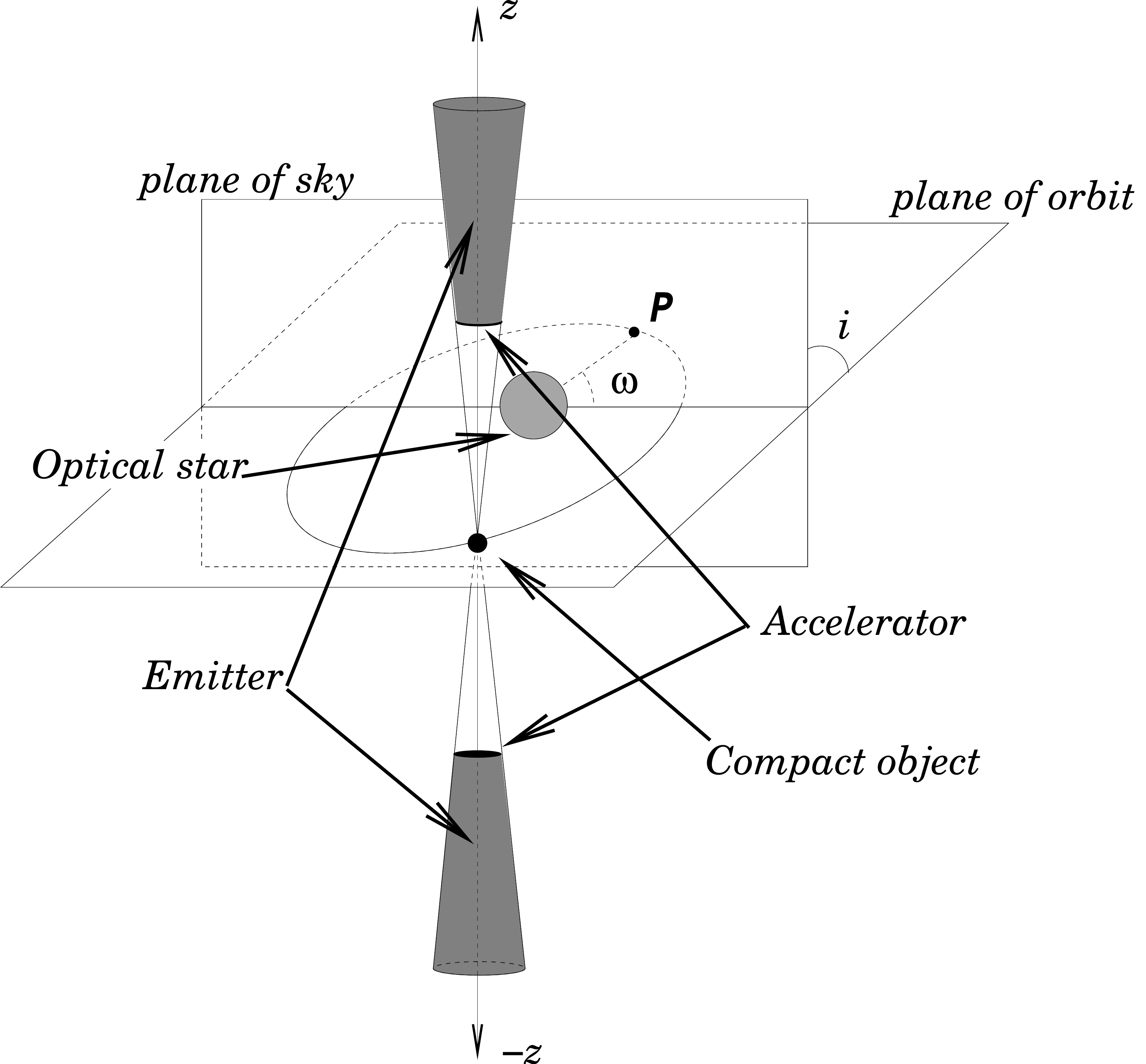}
\caption{Sketch of the system.}
\label{fig:system}
\end{figure}

\section{Physical processes in the system}

\subsection{Radiation and acceleration processes}\label{sec:acceleration}

The gamma-ray spectrum of \source{} extends to very high energies, up to $E_\mathrm{\gamma}\sim 10$~TeV and beyond \citep{hess_ls5039_06}.
Since the parent electrons should have even higher energies, the acceleration of $>10$~TeV electrons requires rather special conditions,
especially if the radiation is formed inside the binary system, where electrons suffer strong IC energy losses \citep{aharonian06}. Assuming for
simplicity { that the gamma-ray emitter is a jet-like
structure}\footnote{The assumption on the jet-like structure here does not necessarily mean adopting the microquasar
scenario (i.e. an accreting system with a jet). In fact, a jet-like structure could also be produced in other types of object, e.g. in binary
pulsar systems \citep{bogovalov07}.}
 perpendicular to the orbital plane, one derives the energy density of the
stellar radiation field in the emitting region: $w_r=L_\star/4 \pi (R_{\rm orb}^2+Z^2) c$, where $L_\star$ is the star 
luminosity and $Z$ is the distance to the compact object.

A sketch of the geometry of the system is shown in Fig.~\ref{fig:system}. The radiation of the companion star in \source{} is characterised
by a black-body spectrum of temperature $kT\approx 3.3$~eV and luminosity $L_\star \approx 7 \cdot 10^{38} \ \rm erg/s$ \citep{casares05}.
Since $Z$ cannot be much larger than the orbital distance (otherwise the periodic component of radiation would be smeared out), the energy
density of the target photons typically varies between $10 \ \rm erg/cm^3$ at $Z\sim10^{13} \ \rm cm$, and  $1000 \ \rm erg/cm^3$ at the base
of the jet around periastron. 

\begin{figure}
\includegraphics[width=0.5\textwidth]{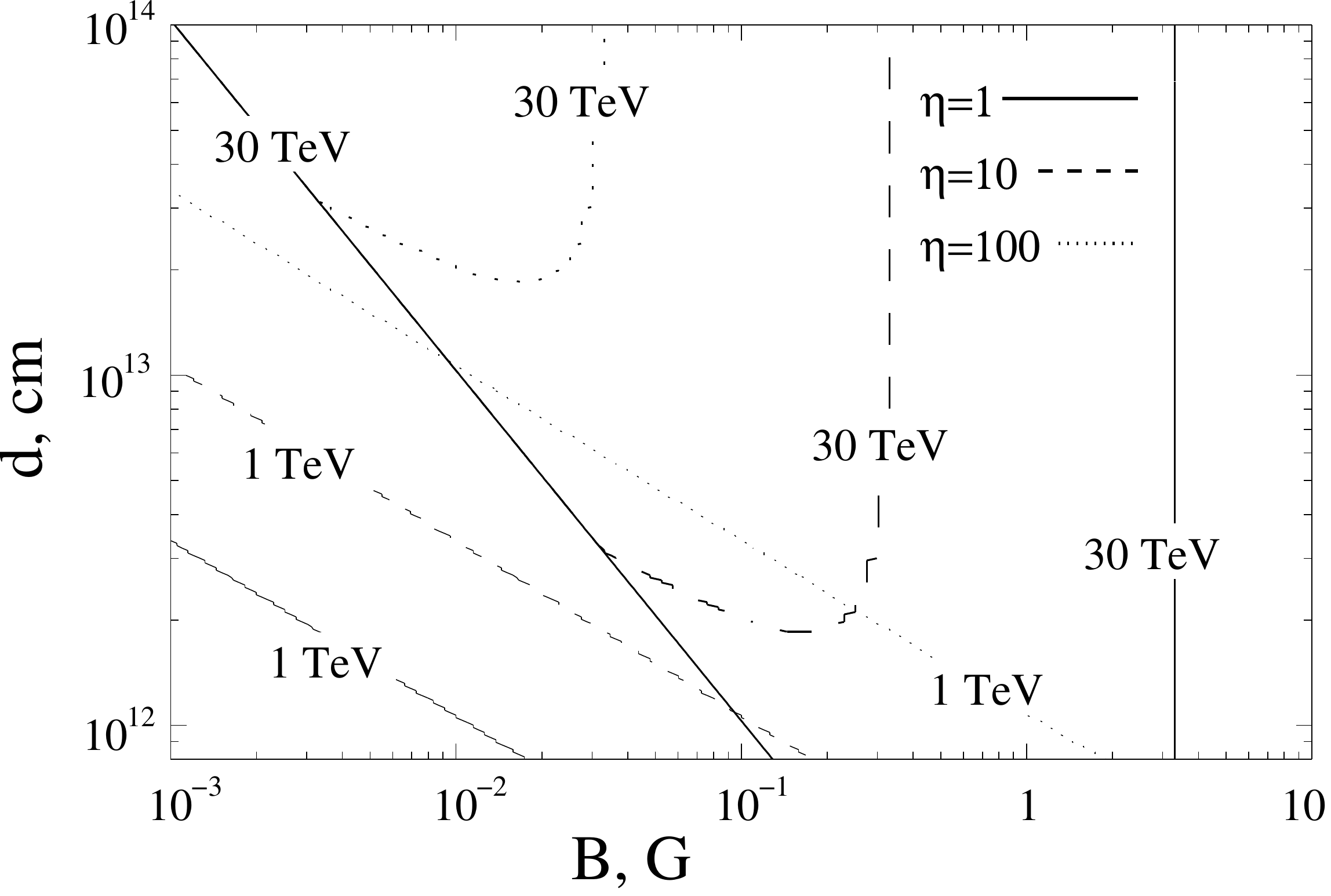}
\caption{Contour plot of { the maximum energy} of accelerated electrons in the B-d plane for different values of the $\eta$ parameter: 
$\eta=1$ (solid lines), $\eta=10$ (dashed lines), $\eta=100$ (dotted lines). Thick and thin 
lines correspond to 30~TeV and 1~TeV electrons, respectively.}
\label{fig:contour}
\end{figure}

{ The IC scattering} of TeV electrons on the starlight photons of average energy $3kT\approx 10$~eV takes place { deep in  the
Klein-Nishina (KN) regime.} The characteristic cooling time of electrons in this regime can be approximated with good
accuracy as \citep{aharonian06}: 
\begin{equation}
t_{\rm KN} \approx 1.7\cdot10^2  
w_0^{-1} E_{e\ \rm TeV}^{0.7} \ {\rm s}
\approx10^3\,d_{\rm13}^{2}E_{e\ \rm TeV}^{0.7}\ {\rm s} \ \,,
\label{ic_cooling}
\end{equation}
where $E_{\rm e~TeV}=E/1\rm TeV$
is the electron energy in TeV units, \mbox{$w_0=w_{\rm r}/100~{\rm erg/cm^{3}}$}, and \mbox{$d_{\rm 13}=\sqrt{R_{\rm orb}^2+Z^2}/10^{13}{\rm cm}$}.  
The maximum acceleration energy of electrons is achieved when the cooling time approaches the acceleration time~($t_{\rm acc}$), 
{ which it is convenient to present} in the following general form:
\begin{equation}
t_{\rm acc}=\eta\, r_{\rm L}/c 
\approx 0.1\, E_{\rm e~TeV}\, B_{\rm 0~G}^{-1}\, \eta \ \rm s\,,\label{accel_time}
\end{equation}
where $r_{\rm L}=E/eB_0$ is the Larmor radius, 
and $B_{\rm 0~G}=B_0/1\ \rm G$ is the strength of { the magnetic field in the accelerator}. 
The parameter $\eta$ characterises { the acceleration efficiency.} 
Generally, $\eta\gg 1$, and only { in so-called extreme accelerators} $\eta \rightarrow 1$ \citep{aharonian02}. 
From the condition  $t_{\rm acc} = t_{\rm KN}$ one obtains:
\begin{equation}
E_{\rm e,max} \simeq  
4\cdot10^{10} \ [B_{\rm 0~G} \eta^{-1}  
w_0^{-1}]^{3.3}~{\rm TeV}.
\label{ic_max_en}
\end{equation}
Formally, { for an extreme accelerator ($\eta\la10$)} and $B_0 \geq 0.3\ {\rm G}$,  the maximum energy of electrons can, under KN IC
energy losses, exceed 100~TeV even for a radiation energy density $w_{\rm r} \simeq 10^3 \ {\rm
erg/cm^3}$. 
However, { for such a magnetic field strength in the accelerator}, synchrotron energy losses
dominate over the Compton losses. The synchrotron cooling time is:
\begin{equation}
t_{\rm sy} \approx 4\cdot10^2 B_{\rm 0~G}^{-2} E_{\rm e~TeV}^{-1} \ \rm s\,,\label{sy_cooling}
\end{equation}
{ thus the condition $t_{\rm acc} = t_{\rm sy}$ gives:
\begin{equation} 
E_{\rm e,max}  \approx 6\cdot10 \, B_{\rm 0~G}^{-1/2} \, \eta^{-1/2}~{\rm TeV} \ .
\label{syn_max_en}
\end{equation}
}

Another fundamental condition, $r_{\rm L}<Z_0$, where $Z_0$ is 
the location of the accelerator and taken as an upper-limit of its linear size ($l_{\rm a~12}=l_{\rm a}/10^{12}$), gives:
\begin{equation} 
E_{\rm e}  < 3\cdot10^2 \, l_{\rm a~12}B_{\rm 0~G}~{\rm TeV}\ \stackrel{l_{\rm a}<Z}{ \Longrightarrow}\ E_{\rm e}  < 3\cdot10^2 \,
Z_{0/12}B_{\rm 0~G}~{\rm TeV}\ ,
\label{size_max_en}
\end{equation}
where $Z_{0/12}=Z_0/10^{12}$.
%
%
\begin{figure}
\includegraphics[width=0.5\textwidth]{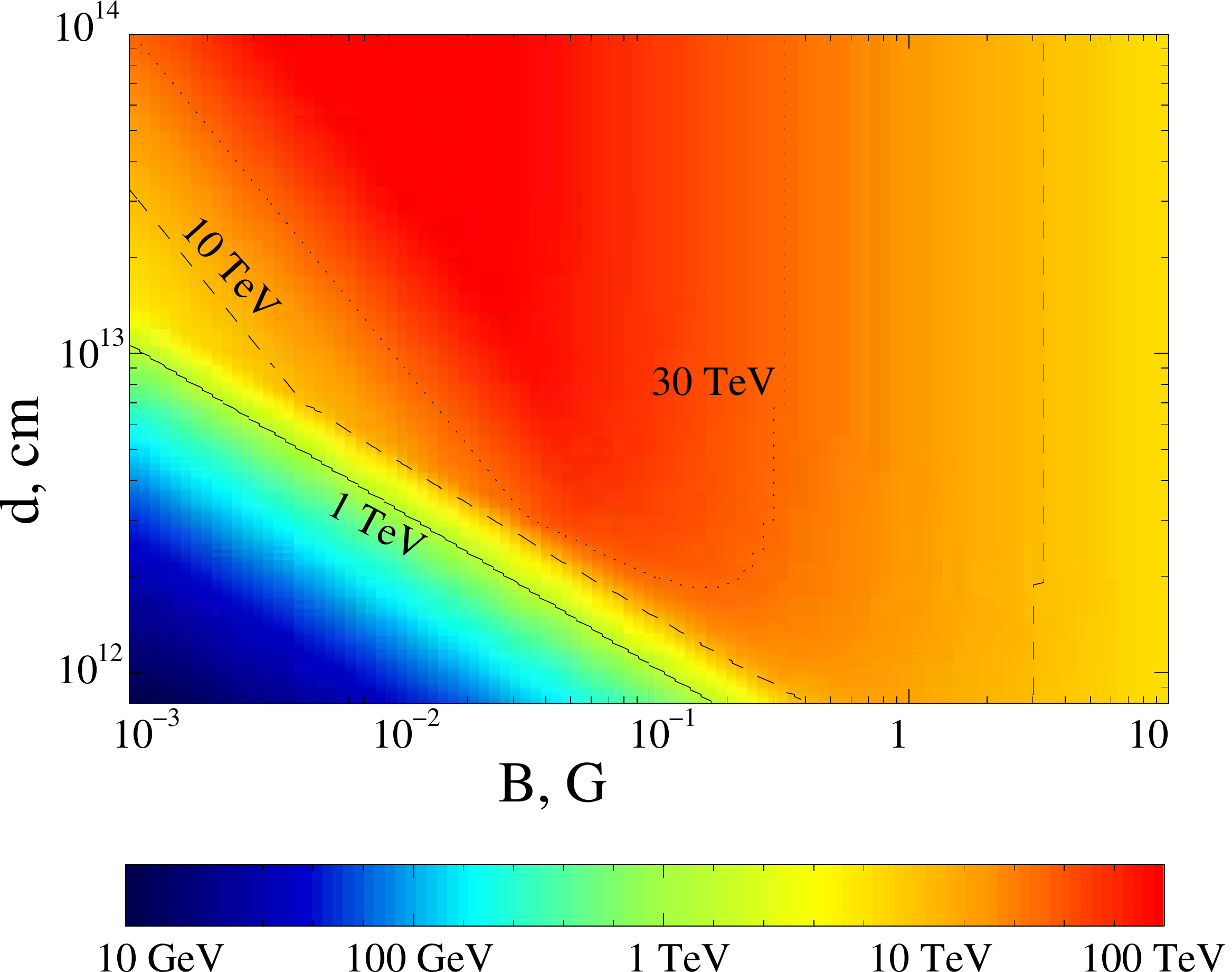}
\caption{The same as in Fig.~\ref{fig:contour} but only for $\eta=10$.}
\label{fig:maxenergy}
\end{figure}

In Fig.~\ref{fig:contour} we show the maximum-energy contour plot 
in the parameter plane $(d,B)$  based on the conditions given by Eqs.~(\ref{ic_max_en}, \ref{syn_max_en}, \ref{size_max_en}). 
The curves correspond to three values of the $\eta$ parameter: 
$\eta=1, 10$ and 100. It is seen that electrons can be accelerated up to 
$\approx 30$~TeV energies deep inside the binary system ($d\simeq 2\cdot 10^{12}$~cm, i.e. $Z_0\ll R_{\rm orb}$) { only in the case of an extreme accelerator} with 
$\eta< 10$. 
It is worth noting that in dense radiation environments
the acceleration efficiency can be significantly enhanced through   
the so-called {\it converter mechanism} \citep{derishev03,stern03}. 
Therefore, quite small values of the acceleration efficiency, i.e. $\eta\sim 1$, cannot be {\it a priori} excluded. 

In Fig.~\ref{fig:maxenergy}, we show the maximum electron energy map in the $(d,B)$ plane { for $\eta=10$}. 
It is seen that for $\eta\ga 10$, electrons can be
accelerated to energies $\ga 30$~TeV only in an environment with $B_0\la 0.3$~G located at $Z_0\ga 10^{12}$~cm. 

The hardness of the \hess{} reported spectra of gamma-rays { allows us to put} 
constrains on the magnetic field strength { in the emitter}. 
Namely, the reported photon indices, ranging { from~$\sim 2-2.5$}, indicate that the magnetic field
energy density in the emitter should be significantly smaller than the target photon energy density.  Indeed, if these gamma-rays are
produced via IC scattering, the electron differential  spectrum must be harder than $\propto E_{\rm e}^{-2}$. For dominant synchrotron
cooling, the  electron energy distribution at high energies is softer than $\propto E_{\rm e}^{-2}$. Even in case of a 
monoenergetic injected
electron spectrum, synchrotron cooling results in a $\propto E_{\rm e}^{-2}$ type electron energy distribution. However, IC energy losses
taking place in the KN regime allow such a harder than $\propto E_{\rm e}^{-2}$ electron energy distribution. 
Thus, hard VHE spectra require $t_{\rm KN}<t_{\rm sy}$, or
\begin{equation}\label{eq:b_hard}
B_{\rm G}<0.6 d_{13}^{-1}E_{\rm e\ TeV}^{-0.85}
\end{equation}
for the magnetic field {  in the emitter}.

\subsection{Propagation of electrons}\label{sec:propagation}

%
\begin{figure}
\includegraphics[width=0.5\textwidth]{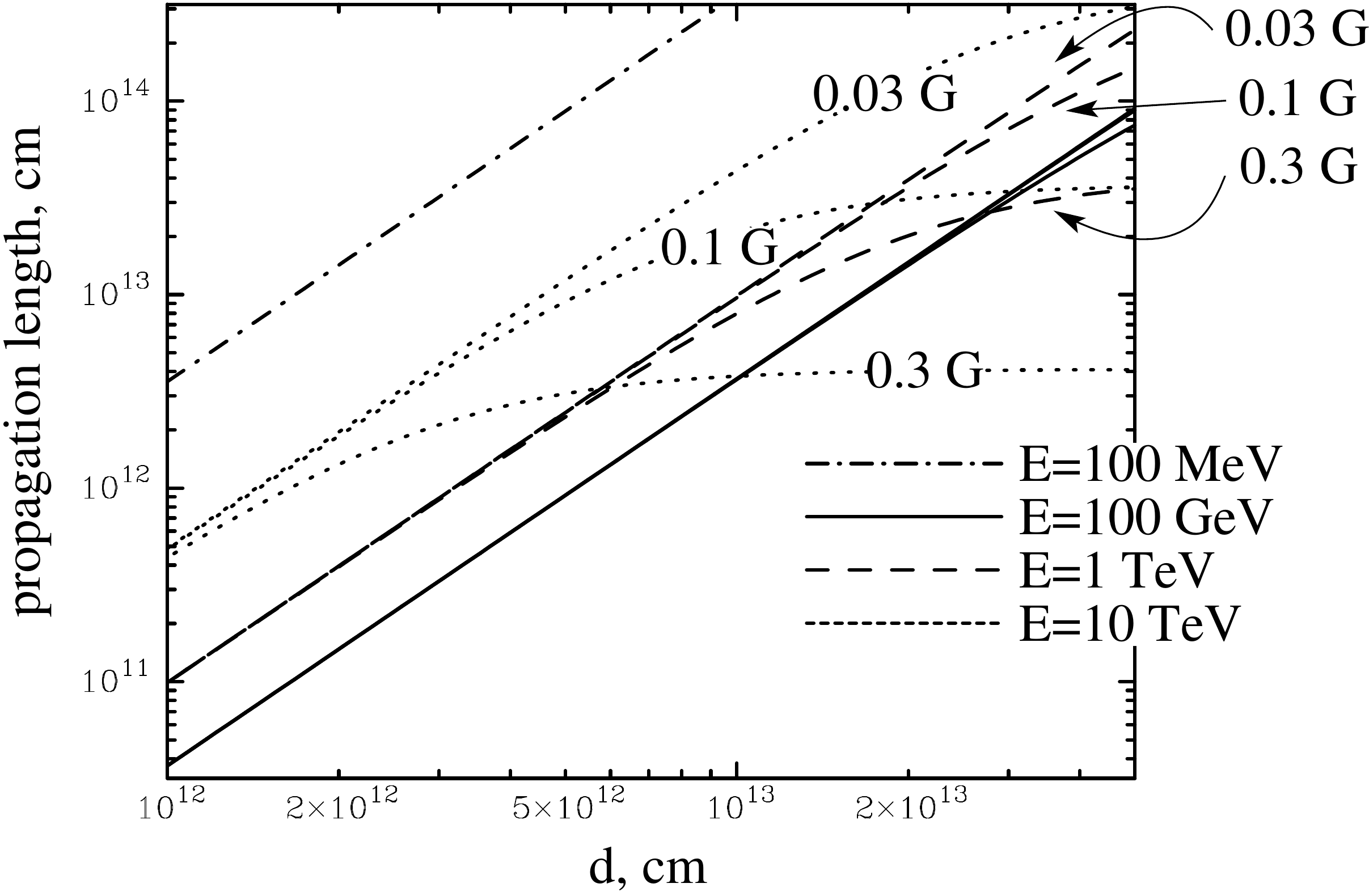}
\caption{Electron propagation length versus location of injection in the system 
under different magnetic fields. Calculations are performed for a black-body 
radiation field with $T=3.8\cdot10^4$~K, 
with a radiation energy density corresponding to that of the star at distance $d$
(i.e. dilution coefficient $\varrho=(R_\star/2d)^2$ with $R_\star=6.5\cdot10^{11}$~cm), 
and for three values of the magnetic field $B=0.03$, $0.1$, $0.3$~G, which are constant along the propagation path. 
The advection velocity was assumed to be $V_{\rm adv}=10^{10}$~cm/s. We note that 
low energy electrons (100~MeV) can propagate very far, whereas 100~GeV electrons have the 
shortest propagation length and, depending on $B$, 10~TeV electrons can propagate farther 
than 1~TeV electrons.}
\label{fig:p-length}
\end{figure}

Relativistic electrons can propagate along the jet. In such a case, their energy cooling proceeds under changing physical
conditions. This can have a strong impact on the resulting gamma-ray radiation. We consider here two possible transport mechanisms:
diffusion and advection along the jet. The propagation length depends on the diffusion coefficient $D$, on the bulk velocity of the jet (or 
advection velocity) $V_{\rm adv}$, and on the radiation cooling time $t_{\rm cool}$. The particle transport is important when the
propagation distance $\Delta Z$  is comparable to the separation between the injection point and the compact object ($Z_0$).
Thus, the particle transport can be described by a dimensionless parameter { $\kappa={\Delta Z/Z_0}$, which is dominated, depending on the propagation regime, by
$\kappa_{\rm diff}=\lambda_{\rm diff}/Z_{0}$ or $\kappa_{\rm adv}=V_{\rm adv}t_{\rm cool}/Z_{0}$.} The diffusion length $\lambda_{\rm diff}=\sqrt{2Dt_{\rm cool}}$,
thus:
\begin{equation}
\kappa_{\rm diff}=0.1\,Z_{0/12}^{-1}E_{\rm e~TeV}^{1/2}B_{\rm G}^{-1/2}\left(t_{\rm cool}\over10^2\rm s\right)^{1/2}\left(D\over D_{\rm Bohm}\right)^{1/2}\,,\label{eq:v_diff}
\end{equation}
where $D_{\rm Bohm}$ is the diffusion coefficient in the Bohm regime, and $E_{\rm e~TeV}$ is the electron energy in TeV units.
Thus, electron diffusion has no impact on the TeV radiation, unless the diffusion is far from the Bohm regime 
($D\gg D_{\rm Bohm}$) or the magnetic field is very small in the emitter. 
For advection we obtain:
\begin{equation}
\kappa_{\rm IC}=10{V_{\rm adv}\over 10^{10}{\rm cm/s}} d_{13}^2Z_{0/12}^{-1}E_{\rm e~TeV}^{0.7}\,,
\end{equation}
when energy losses are dominated by IC scattering, and
\begin{equation}
\kappa_{\rm syn}=4{V_{\rm adv}\over10^{10}{\rm cm/s}}  Z_{0/12}^{-1}E_{\rm e~TeV}^{-1}B_{\rm G}^{-2}\,,
\end{equation}
for dominant synchrotron energy losses. 
Therefore, for a mildly relativistic outflow ($V_{\rm adv}\sim 10^{10}$~cm/s), $\kappa$ can easily exceed 1 
under dominant KN IC energy losses. 
In Fig.~\ref{fig:p-length}, the results of the electron advection length calculations are shown for four different
energies $E_{\rm e}=100$~MeV, $100$~GeV, 1~TeV, and $10$~TeV. Here we consider a $V_{\rm
adv}=10^{10}$~cm/s. Three constant values of the magnetic field were considered:
$B=0.03$, $0.1$, and $0.3$~G. 
The radiation field was assumed to be black-body with $T=3.8\cdot10^4$~K, and energy density corresponding to that 
of the star at distance $d$ (i.e. dilution coefficient
$\varrho=(R_{\star}/2d)^2$, where $R_{\star}=6.5\cdot10^{11}$~cm). As seen in Fig.~\ref{fig:p-length}, electrons with energy
$E_{\rm e}\sim 100$~GeV have the shortest advection length. Due to cooling in the Thomson regime ($t_{\rm T}\propto E_{\rm
e}^{-1}$), electrons with smaller energy have a longer propagation length.
In case of higher electron energy ($E_{\rm e}\ga100$~GeV), IC cooling proceeds in
KN regime, leading to an increase of the propagation length with the growth of the electron energy unless synchrotron energy losses
start to dominate. {  All this is clearly seen} in Fig.~\ref{fig:p-length}.

As discussed above, the  parameter $\kappa$ {\rm varies over a rather wide range} depending on the electron energy and the accelerator-companion
star separation distance. Thus, high energy electrons can propagate far away from the acceleration site. In the next section we show
that this effect can significantly change the gamma-gamma optical depth for the produced VHE radiation and the IC scattering angle between 
electrons and target photons.

\subsection{IC and gamma-gamma absorption}\label{sec:angles}

\begin{figure}
\includegraphics[width=0.5\textwidth,angle=0]{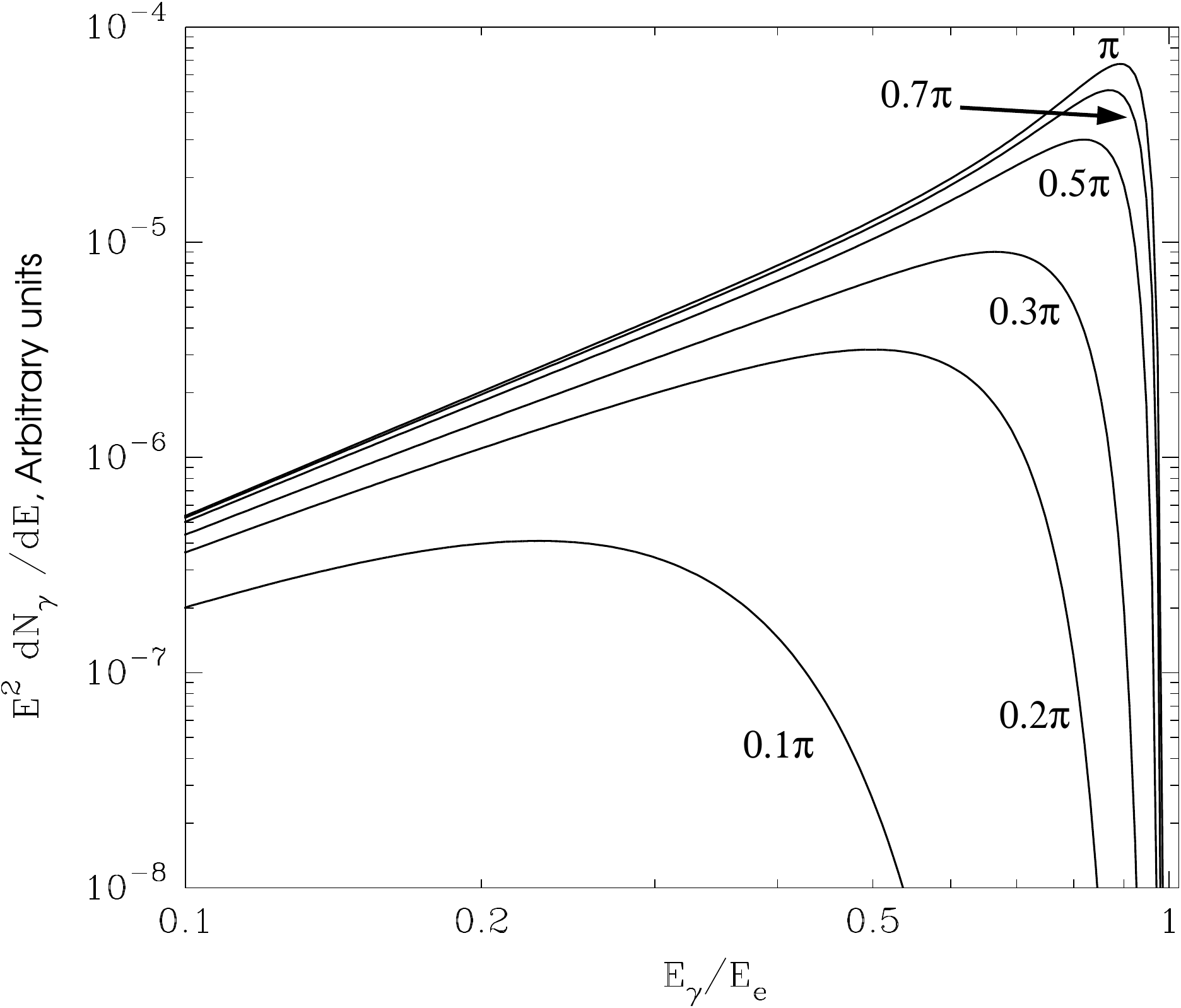}
\caption{Dependence of the IC energy spectrum on the interaction angle: 
$\theta_{\rm IC}=0.1\pi,0.2\pi,0.3\pi,0.5\pi,0.7\pi,\pi$. It is assumed that monoenergetic electrons 
($E_\mathrm{e}=0.1$~TeV) interact with black-body photons ($T=3.8\cdot10^4$~K).}
\label{fig:icangle}
\end{figure}
%

\begin{figure}
\includegraphics[width=0.5\textwidth,angle=0]{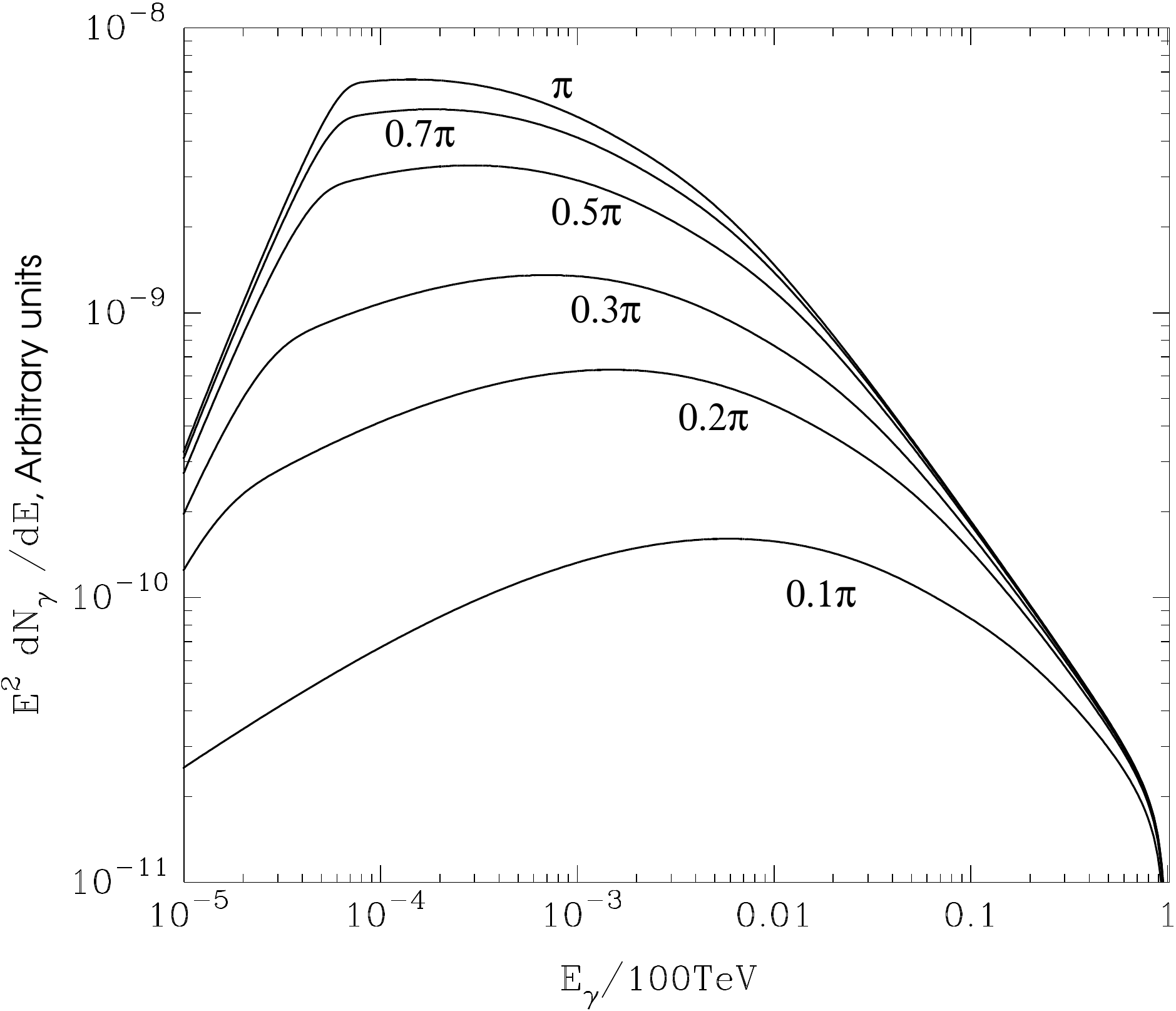}
\caption{Angular dependence of the IC cross-section. The result of the interaction of a powerlaw electron distribution with a black-body 
distribution of target photons ($T=3.8\cdot10^4$~K) is shown for the following values of the interaction angle:  
$\theta_{\rm IC}=0.1\pi,0.2\pi,0.3\pi,0.5\pi,0.7\pi,\pi$.}
\label{fig:icangle_pl}
\end{figure}

\begin{figure}
\includegraphics[height=0.5\textwidth,angle=270]{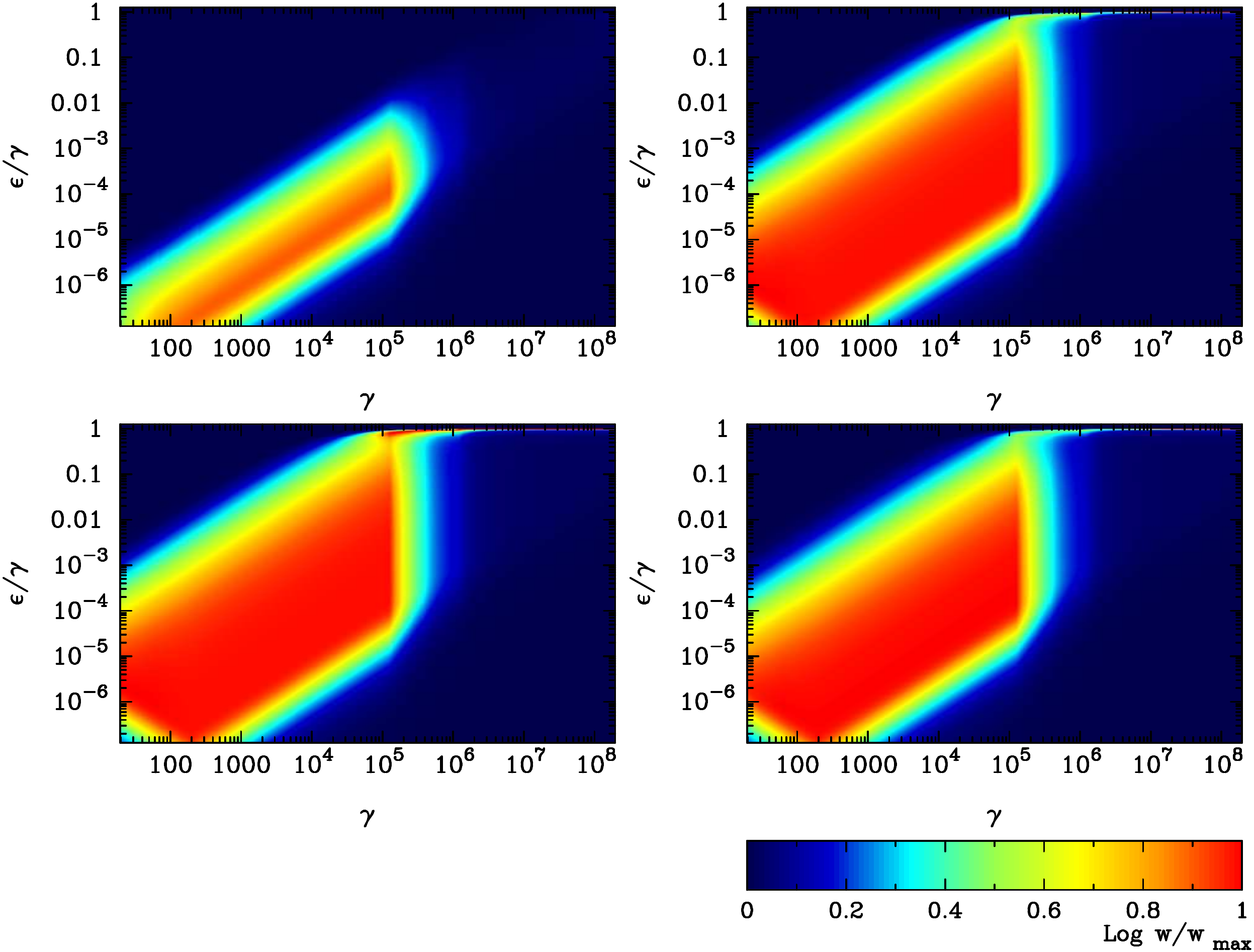}
\caption{ A 2D representation in the ($\gamma$,$\epsilon/\gamma$) plane of the IC interaction probability for different $\theta=5^\circ$
(upper-left panel), $90^\circ$ (upper-right panel), $175^\circ$ (lower-left panel). The same is shown for the \textit{angle averaged} IC interaction probability  (lower-right panel).}
\label{fig:icbig}
\end{figure}

Because of the strong anisotropy and inhomogeneity of the target photon distribution, IC scattering and gamma-gamma absorption depend on the
electron location and momentum direction. Regarding IC scattering, the interaction angle $\theta_{\rm IC}$ has a strong impact on the
formation of radiation as shown in the past by several authors in different astrophysical scenarios (e.g. \cite{bogovalov00,khangulyan2,khangulyan_mor,dermer06}). In
Fig.~\ref{fig:icangle}, we show the IC spectral energy distribution (SED) produced by a monoenergetic { distribution of electrons 
of} $E_\mathrm{e} = 0.1$~TeV 
interacting { with photons} with a black-body distribution ($T = 3.8\cdot10^4$~ K) at different angles $\theta_{\rm IC}$ ranging from
$0.1\pi$ to $\pi$. It is seen that, depending on $\theta_{\rm IC}$, for electron energies $E_\mathrm{e}\ga m^2c^4/kT$, the IC scattering can
proceed either in the KN ($E_\mathrm{\gamma}/E_\mathrm{e}\sim 1$) or in the Thomson ($E_\mathrm{\gamma}/E_\mathrm{e}\ll 1$) regimes. In
{ In Fig.~\ref{fig:icangle_pl}, we show the computed SED corresponding to the IC scattering between electrons with a powerlaw energy
distribution and photons with a black-body distribution ($T = 3.8\cdot10^4$~K). Results are shown for different $\theta_{\rm IC}$
ranging from $0.1\pi$ to $\pi$. Since the interaction angle strongly varies along the orbit and with $Z$,  the final gamma-ray
spectra significantly depend on the orbital phase and location of the emitter.}
Fig.~\ref{fig:icbig}, we show a two-dimensional representation of the IC scattering probability in the plane 
($\gamma$,$\epsilon/\gamma$; where $\gamma$ is electron Lorentz factor and $\epsilon$ is the outgoing photon energy) for
three different $\theta_{\rm IC}=5^\circ$, $90^\circ$ and $175^\circ$, which are shown altogether with the average cross-section case.
Depending on the electron distribution function, the IC anisotropy could have a 
different impact on the gamma-ray spectrum. In the case of a
monoenergetic or narrow electron distribution (e.g. $\delta$-function) { around 0.1~TeV ($\sim m^2c^4/kT$)}
the IC scattering at high interaction angles results in higher fluxes
and harder spectral shapes (see Fig.~\ref{fig:icangle}). For broad (e.g. power-law) electron distributions, IC scattering at small
interaction angles leads to a rather low flux, but the spectral shape is harder than for large scattering angles. The hard spectral shape
is due to  a transition from the Thomson, in which the scattering is sensitive to the interaction angle, to the KN regime,
in which the angular dependence is weak. We note 
that the smaller the interaction angle, the larger the transition energy.

\begin{figure}
\includegraphics[width=0.5\textwidth,angle=0]{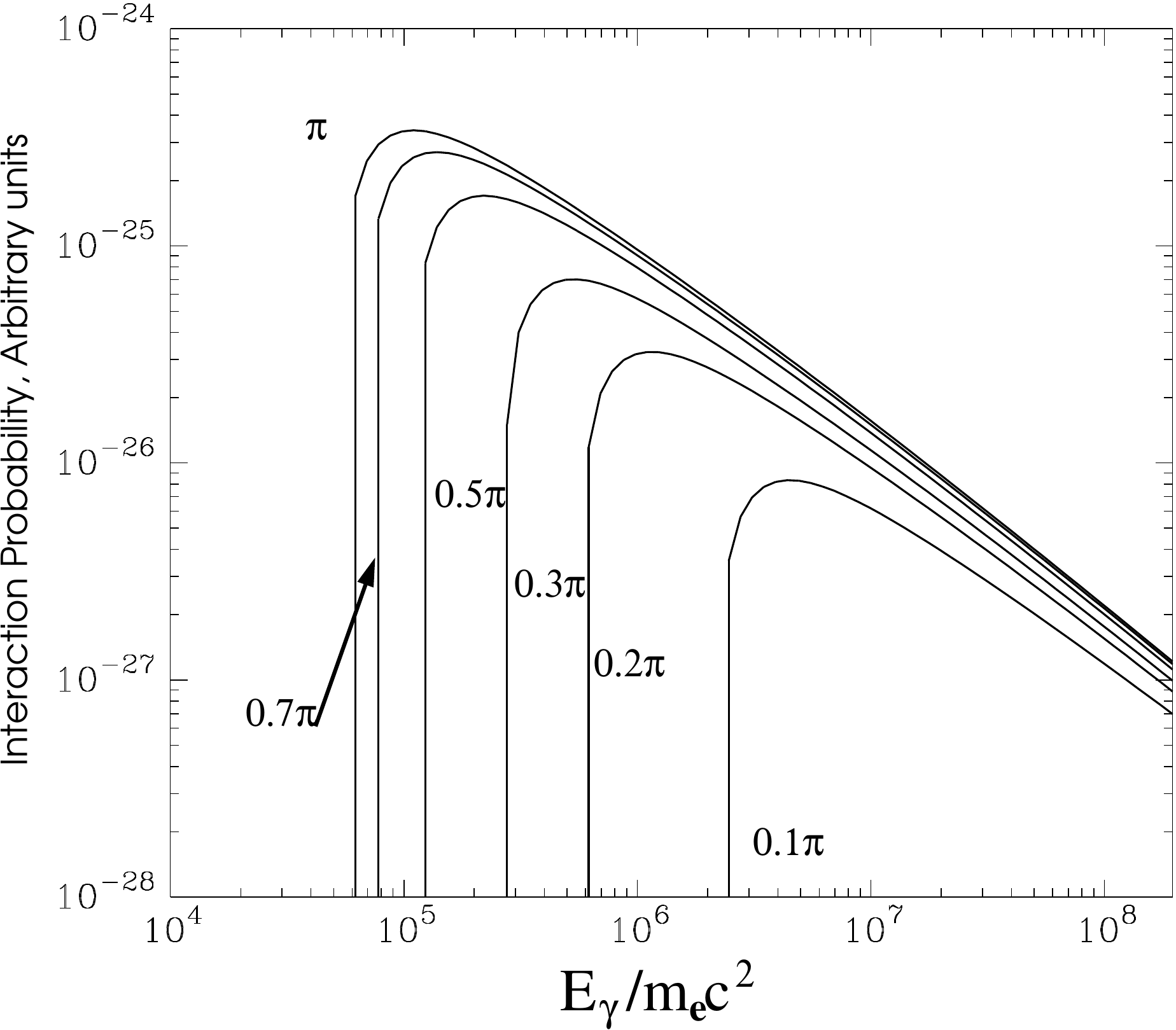}
\caption{The probability of gamma-gamma absorption for a monoenergetic distribution of target photons ($\epsilon=10$~eV) for the 
following values of the interaction angle: $\theta_{\rm \gamma\gamma}=0.1\pi,0.2\pi,0.3\pi,0.5\pi,0.7\pi,\pi$.}
\label{fig:ggangle}
\end{figure}

The gamma-ray absorption in close binaries was studied by, e.g., \cite{moskalenko94}, \cite{boettcher05} and \cite{dubus05}. In general, the
gamma-gamma optical depth depends on the distance $d$ between the emitting electrons and the companion star, and on the scattering angle (assuming a point-like source of target photons)
$\theta_{\rm IC}$:
\begin{eqnarray}
 \tau(d,\theta_{\rm IC},E_\gamma)=\int\limits_{\rm line\ of\ sight}{\rm d}l\ (1-\cos{\theta_{\gamma\gamma}})\nonumber\\
\times  \int{\rm d}\epsilon\ \sigma_{\rm pp}(\epsilon
 E_{\gamma}(1-\cos{\theta_{\gamma\gamma}}))n_{\rm ph}(\epsilon).
 \label{eq:optical_depth}
\end{eqnarray} 
where $\theta_{\gamma\gamma}=\theta_{\gamma\gamma}(l)$ is the angle between the directions of the gamma-ray and the stellar photon at distance $l$ 
from {  the emitting point}. 
The absorption probability strongly depends on
the interaction angle. This is shown in Fig.~\ref{fig:ggangle}, where the probability of gamma-gamma absorption on a monoenergetic
distribution of target photons ($\epsilon=10$~eV) is plotted for the interaction angle $\theta_{\gamma\gamma}=0.1\pi, 0.2\pi, 0.3\pi, 0.5\pi, 0.7\pi,
\pi$. Concerning the $d$-dependence of the optical depth, in the point-like approximation for the geometry of the target photon
source, { the gamma-gamma optical depths for arbitrary $d_0$ and 
$d$ can be scaled as:}
\begin{equation}
 \tau(d,\theta_{\rm IC},E_\gamma)={d_0\over d}\tau(d_0,\theta_{\rm IC},E_\gamma).
\end{equation}
%

From previous considerations, the smaller $\theta_{\rm IC}$ the harder the produced radiation. In addition, since $\theta_{\rm IC}\ll 1$
implies also $\theta_{\gamma\gamma}\ll 1$ in Eq.(\ref{eq:optical_depth}), 
the { threshold energy} for gamma-gamma absorption significantly increases up to 
\mbox{$E_{\rm th}\gg 100$~GeV}.
Both effects result in hard and unabsorbed gamma-ray spectrum above 1~TeV. 

%

\subsection{Radiation of secondary electrons}\label{sec:secondary}

\begin{figure}
\includegraphics[height=0.5\textwidth,angle=270]{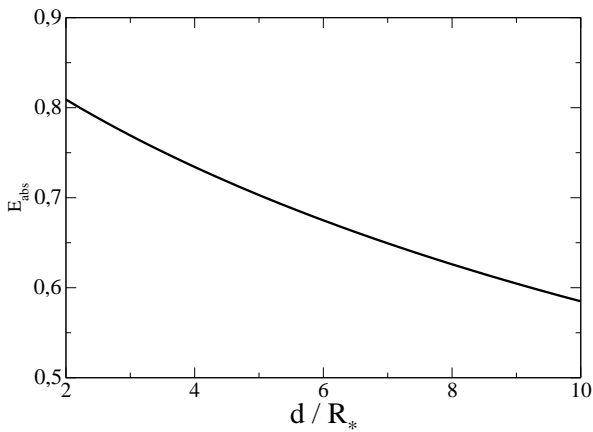}
\caption{{ Fraction of the primary gamma-ray energy emitted in the range $0.1$~--~$10$~TeV ($E_{\rm abs}$) which was absorbed as a function of the distance to the star in stellar radius units.  We assumed a power-low distribution of primary gamma-rays ($\propto E_\mathrm{\gamma}^{-2}$).}}
\label{fig:absorp_en}
\end{figure}

The interaction of gamma-rays with the stellar radiation field produces a population of secondary electron/positron pairs
(secondary electrons hereafter) inside the binary system. A significant fraction { of the primary gamma-ray energy} can be transfered to these particles. In
Fig.~\ref{fig:absorp_en}, we show an example of the fraction of gamma-ray energy absorbed in the radiation field of the star. Assuming an
isotropic point-like emitter located at distance $d$ from the optical star, we calculate the absorbed fraction of primary gamma-ray energy in
the energy range $0.1-10$~TeV (for a $E_{\gamma}^{-2}$ spectrum of primaries). One can see that, even for an emitter located rather far from
the star, radiation of secondaries can significantly contribute to the observed emission from the source. The secondary electrons radiate via
the synchrotron or IC channels. The ratio of these fluxes depends on the energy band, the magnetic field strength and the distance to the
companion star. This effect may lead to the formation of an electromagnetic cascade as long as the energy of electrons and gamma-rays 
exceeds a certain energy:
\begin{equation}
E > {m^2c^4\over \epsilon_{\rm ph}}=30\left({\epsilon_{\rm ph}\over10{\rm eV}}\right)^{-1}\,{\rm GeV}\ .
\end{equation}
For the effective development of the cascade, IC energy losses should dominate over synchrotron energy losses. 
From Eqs.~(\ref{ic_cooling}) and (\ref{sy_cooling}) one obtains:
\begin{equation}
E_{\rm e~TeV}^{1.7}<2\cdot10^{-2}w_{\rm r}B_{\rm a}^{-2}\,,
\end{equation}
or
\begin{equation}
E_{\rm e}< 1 \left(w_{\rm r}\over10^3w_B\right)^{0.6}\, {\rm TeV}\ ,
\end{equation}
{ where $w_B=B_{\rm a}^2/8\pi$ is the energy density of the ambient magnetic field in the system,
which is distinguished here from the magnetic field inside the emitter.} 
This relation allows us to introduce a critical value of the magnetic field in the binary system:
\begin{equation}
B_{\rm a~c}=10\left(L_{\rm surf}\over 7\cdot10^{38} {\rm  erg\over s}\right)^{1/2}\left(R\over R_{\star}\right)^{-1}{\rm G}\ , 
\label{eq:b_crit}
\end{equation}
where $R$ is the distance between the gamma-ray absorption point and the star.
If the magnetic field is stronger than $B_{\rm a~c}$, the electromagnetic cascade will not affect the energy range 
$E_{\rm e}>1$~TeV, implying a purely absorbed spectrum for $E_{\gamma}>1$~TeV.
%
%

Although the magnetic field within the system is unknown, the magnetic { field at the surface} of O type stars can be typically 
$\approx 100$~G~--~$1$~kG 
(see e.g. \cite{usov92,donati02}). Taking into account the value of $B_{\rm a~c}$ (see Eq.~(\ref{eq:b_crit})), 
efficient cascading at $\ga
1$~TeV seems unlikely. For such magnetic fields, synchrotron emission is the main radiation channel at these energies, and electromagnetic 
cascading will be suppressed.

\section{Model description}

\begin{figure}
\includegraphics[width=0.5\textwidth]{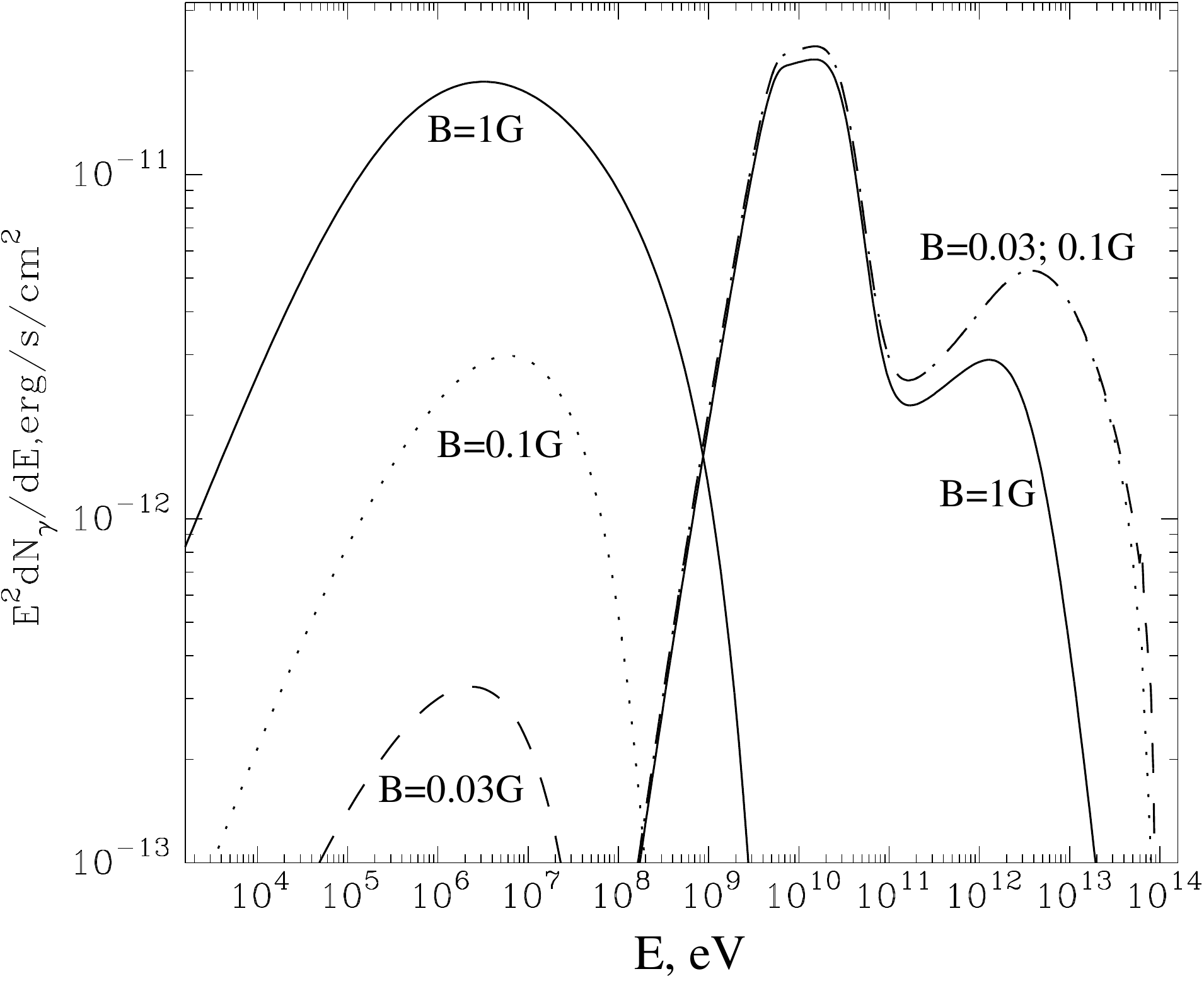}
\caption{Calculated SED for three different $B_0=0.03$, $0.1$ and $1$~G. 
The rest of parameters have been fixed to the next values: $Z_0=2\cdot10^{12}$~cm, $\eta=1$,
$V_{\rm adv}=10^{10}$~cm/s, and $i=25^\circ$. Gamma-ray absorption is included.}
\label{fig:sed_b0}
\end{figure}

In this work, we adopt a model for gamma-ray emission in \source{} in which emission is produced in a jet-like structure perpendicular to the orbital
plane. Also, we assume that the acceleration takes place in a compact region located in the jet at $Z_0$ ({ although the emitting region} could be much
larger because of particle advection) and $Z_0$ does not depend on the orbital phase. In Fig.~\ref{fig:system}, 
the accelerator and the emitter are shown as a black and a { grey region}, respectively. 

The acceleration/cooling time-scales in the system are much shorter than the orbital period. Therefore, the electron distribution function $n$ can be
described by a steady-state equation. In the case of an 
one-dimensional jet, $n(\gamma,Z)$ is determined by the following equation: 
\begin{equation}
V_{\rm adv}{\partial n\over \partial z}+{\partial \dot{\gamma}(\gamma,Z)n\over \partial \gamma}=V_{\rm adv}q(\gamma)\delta(Z-Z_0)\,,
\label{eq:density}
\end{equation}
where the $V_{\rm adv}$ is the jet bulk velocity; $\dot{\gamma}(\gamma,Z)$ is the electron energy loss rate, which
accounts for IC and synchrotron mechanisms and depends both on the electron energy and position in the jet; 
$Z_0$ is { the location of the accelerator}; and $q(\gamma)$ is the injection electron spectrum.
We note that $q(\gamma)$ and $\dot\gamma$ are slow functions of time, i.e. they 
vary adiabatically along the orbit. { Because of a lack of knowledge} concerning the
acceleration processes occurring in \source{}, we adopt a phenomenological function for $q(\gamma)$. Namely, we
assume a power-law function with an exponential high-energy cutoff:
\begin{equation}
q(\gamma)=Q_{\rm0}\ \gamma^{-\alpha}\exp\{-\gamma mc^2/E_{\rm 0}\}\ .
\end{equation}
$E_{\rm 0}$ is the electron cutoff energy determined from the balance between acceleration and energy loss rates 
(see Sec.~\ref{sec:acceleration}); $\alpha$ is fixed to 2, which is a common value for injected spectra of non-thermal electrons, and $Q_0$ is a constant. 

We have numerically solved Eq.~(\ref{eq:density}) taking into account IC and synchrotron radiation processes. We neglect at this stage
ionization or Bremsstrahlung losses because they are not relevant in the present context.  We
assume that the main targets for IC scattering are provided by the star, and that the magnetic field decreases with the distance as
$B=B_0(Z/Z_0)^{-1}$. In order to accelerate electrons to energies $\ga$10~TeV 
the magnetic field should be within $B_0\sim 0.01-1$~G, depending on
the acceleration rate and location (see Fig.~\ref{fig:contour}). 

To study the impact of the magnetic field on the IC and synchrotron radiation components, three different $B_0=0.03$, $0.1$ and $1$~G have
been adopted. The rest of relevant parameters have been fixed to the values: $Z_0=2\cdot10^{12}$~cm, 
$\eta=1$, $V_{\rm adv}=10^{10}$~cm/s, and 
$i=25^\circ$. The corresponding broadband SED is shown in Fig.~\ref{fig:sed_b0}. We note a very hard synchrotron spectra in the interval
$1$~keV~--~$1$~MeV (with a spectral energy index $\sim 0.5$).  This is explained by the hardening of electron energy distribution due to the
radiative cooling in KN regime \citep{khangulyan2,moderski}. Since for magnetic fields $B<1$~G the energy losses are dominated by IC
scattering, the synchrotron radiation flux is significantly less than the IC gamma-ray flux, 
and below also the observed fluxes at X-rays \citep{bosch07}.

%
\begin{figure}
\includegraphics[height=0.5\textwidth,angle=270]{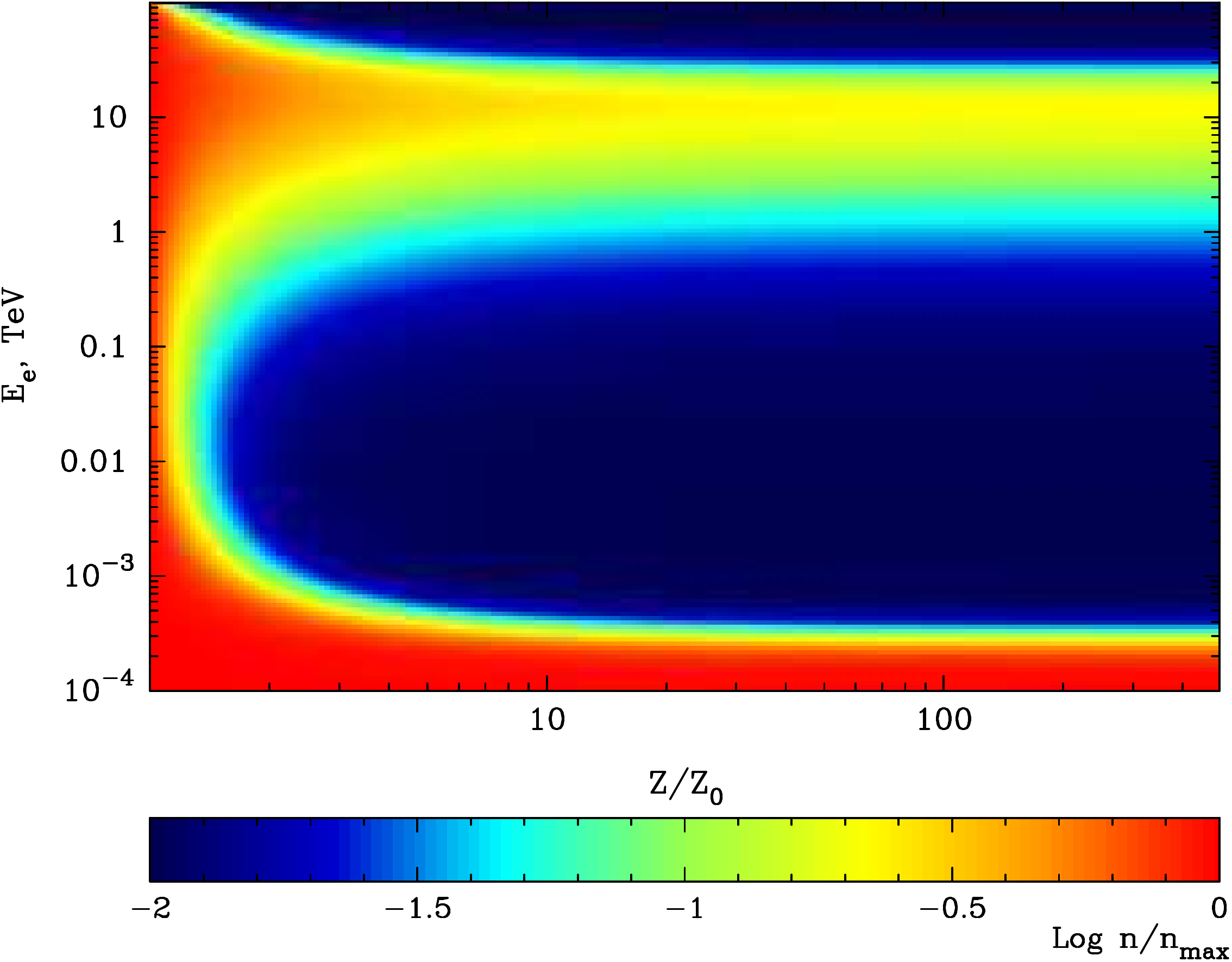}
\caption{The distribution of electrons along the jet for different energies. 
$B=0.2(Z/Z_0)^{-1}$~G, $Z_0=2\cdot 10^{12}$~cm, and $V_{\rm adv}=10^{10}$~cm/s.}
\label{fig:prop_2d}
\end{figure}

{ It is worth discussing} the importance of the electron propagation, { already commented on} in 
Sec.~\ref{sec:propagation}. Namely, we expect the most efficient propagation of electrons at very low (Thomson regime) 
and at very high (deep KN regime) energies as long as IC energy losses dominate over synchrotron ones. In Fig.~\ref{fig:prop_2d}, we show the distribution of electrons along the jet. 
It is seen that, the higher the energy, the larger the
propagation length of electrons. 
{ This is simply the result of the KN effect. However, for the assumed magnetic field, this tendency terminates
around 30~TeV, when the IC energy losses are overcome by the synchrotron ones.}
We note also that, below $E_{\rm e} \la 10$~GeV, the lower the energy, the longer
the propagation length of electrons, because in this energy interval the IC scattering occurs in the Thompson regime.  

\begin{figure}
\includegraphics[width=0.5\textwidth]{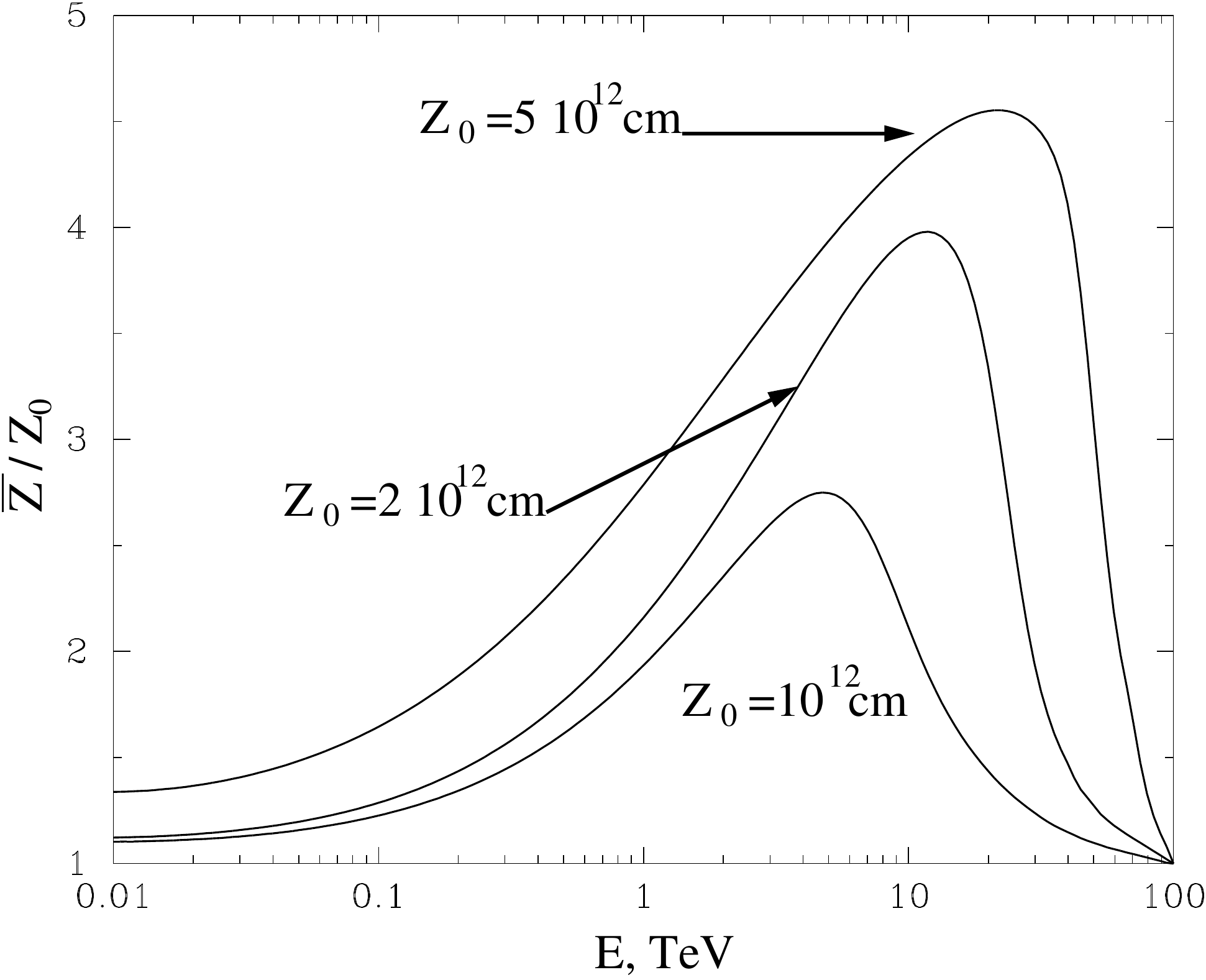}
\caption{The electron $1/d^2$-weighted average location for three different $Z_0$, i.e. $1\cdot10^{12}$, $2\cdot10^{12}$, 
and $5\cdot10^{12}$~cm, adopting $V_{\rm adv}=10^{10}$~cm/s and $B=0.4 (z/10^{12}{\rm cm})^{-1}$~G.}
\label{fig:prop}
\end{figure}

In fact, electrons at large $Z$ will interact with weaker magnetic and 
stellar radiation fields, thus it is convenient to introduce a parameter which characterises the {\it effective} electron average location:
\begin{equation}
 \bar{Z}(E_\mathrm{e})={\int\limits_{Z_0}^\infty (Zn(Z,E_{\rm e})/d^2)\mathrm{d}Z \over \int\limits_{Z_0}^\infty (n(Z,E_{\rm e})/d^2)\mathrm{d}Z}.
 \label{aver}
\end{equation}
The  
$Z_0$-dependence of this {\it effective} electron average location is shown in
Fig.~\ref{fig:prop}, where the $1/d^2$-weighted average locations 
of the electrons of different energy are shown for three different 
$Z_0=1\cdot10^{12}$, $2\cdot10^{12}$, 
and $5\cdot10^{12}$~cm. $B=0.4(Z/10^{12}{\rm cm})^{-1}$~G and $V_{\rm adv}=10^{10}$~cm/s are adopted for all
three cases\footnote{We recall (see Eq.~\ref{eq:v_diff}) that diffusion is negligible for these parameter values.}. The $1/d^2$-weight comes from the fact that the stellar radiation field decreases as $1/d^2$, giving 
an idea on where electrons of a certain energy will radiate the most via IC.
We see that $\bar{Z}(E_\mathrm{e})$ 
significantly depends not only on the energy but also on the location of the accelerator. 

\begin{figure}
\includegraphics[width=0.5\textwidth]{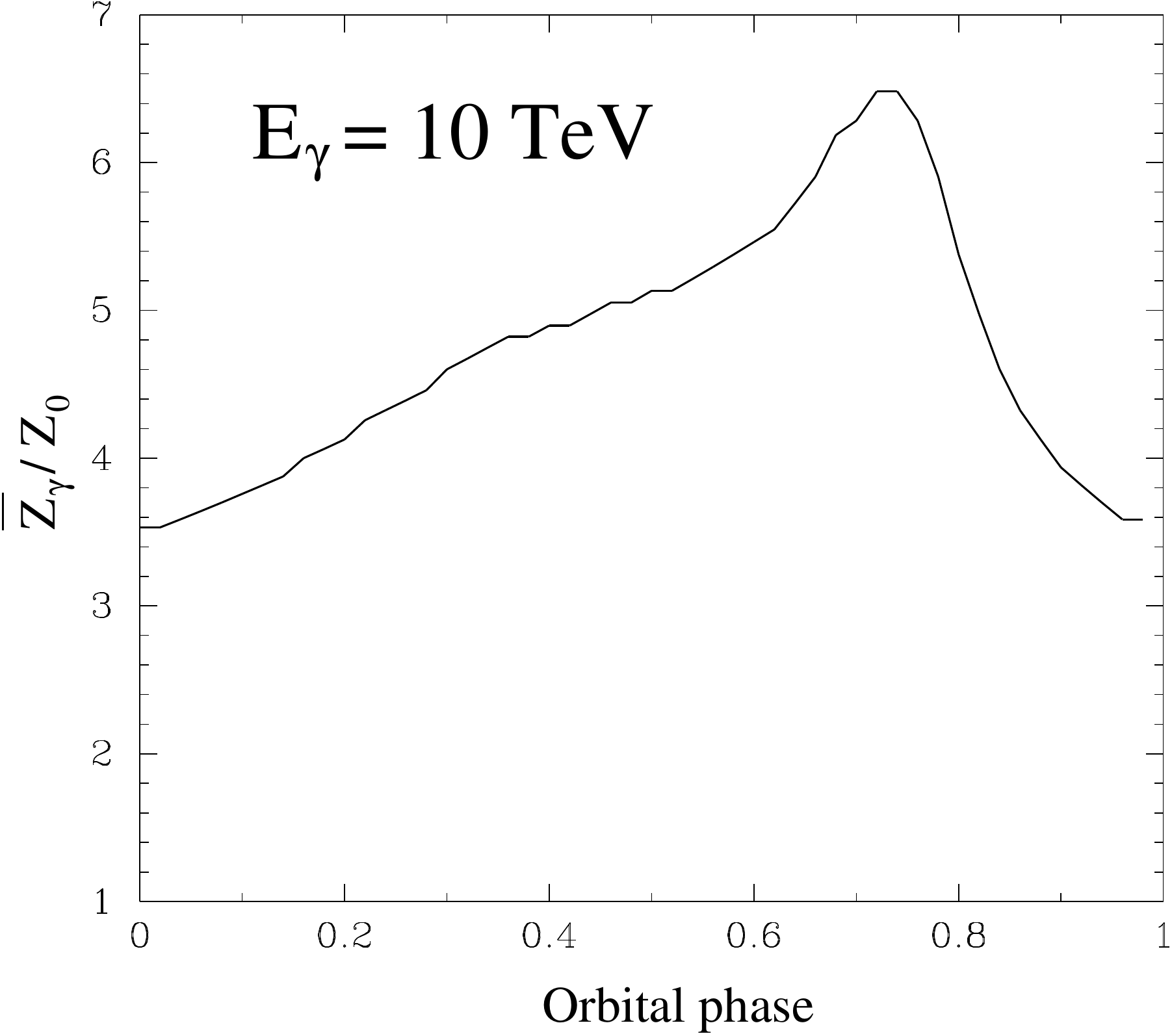}
\caption{The averaged location for the production of 10~TeV gamma-rays produced in the jet. 
$Z_0=2\cdot10^{12}$~cm and $i=25^\circ$ are adopted. This shows the importance of the IC interaction angle
for the resulting gamma-ray production.}
\label{fig:el_prop}
\end{figure}

\begin{figure}
\includegraphics[width=0.5\textwidth]{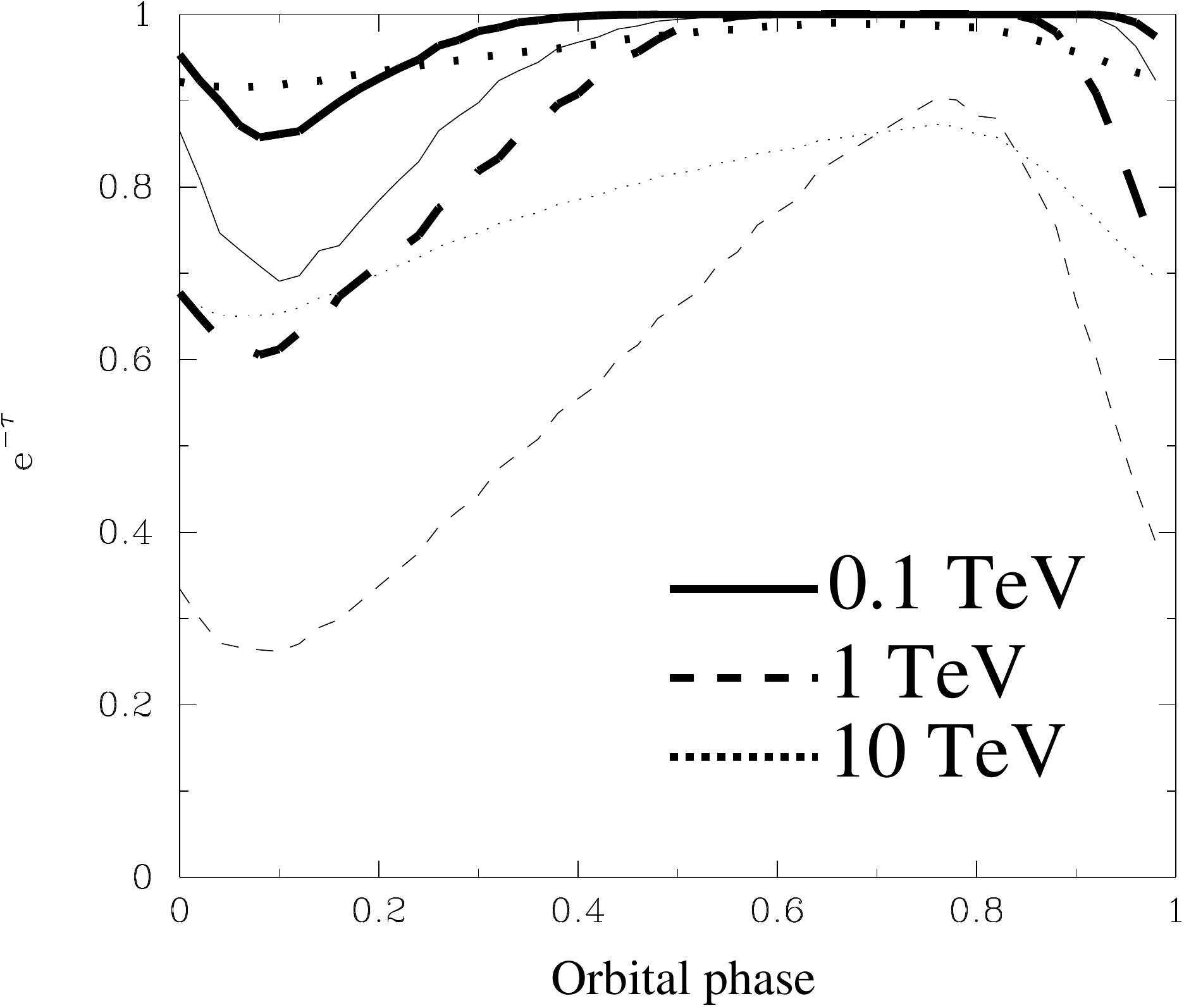}
\caption{The gamma-gamma absorption factor, $\exp{(-\tau)}$, corresponding to three different energies for the gamma-ray produced in the jet: 
$0.1$ (solid lines) $1$ (dashed lines), and $10$~TeV (dotted lines).
Thin lines represent the absorption factor of gamma-rays produced in the accelerator, 
at $Z_0=2\cdot10^{12}$~cm. The absorption factors relevant for 
the average location of the gamma-ray emitter are shown with thick lines. 
The inclination angle is assumed to be $i=25^\circ$. This shows the importance of the gamma-gamma interaction
angle for the gamma-ray absorption.}
\label{fig:absorption}
\end{figure}

This rather unusual 
energy-dependence of the propagation of electrons in the jet can be reflected in the features of the electron radiation. We
note, however, that the spatial distribution of the gamma-rays along the jet does not simply repeat the distribution of electrons. Two
additional effects, related to the interaction angle in the IC scattering and gamma-gamma absorption, lead to a significant modification 
of the
{ gamma-ray effective distribution} along the jet. These effects are shown in Figs.~\ref{fig:el_prop} and \ref{fig:absorption}, respectively. 
%

\begin{figure}
\includegraphics[width=0.5\textwidth]{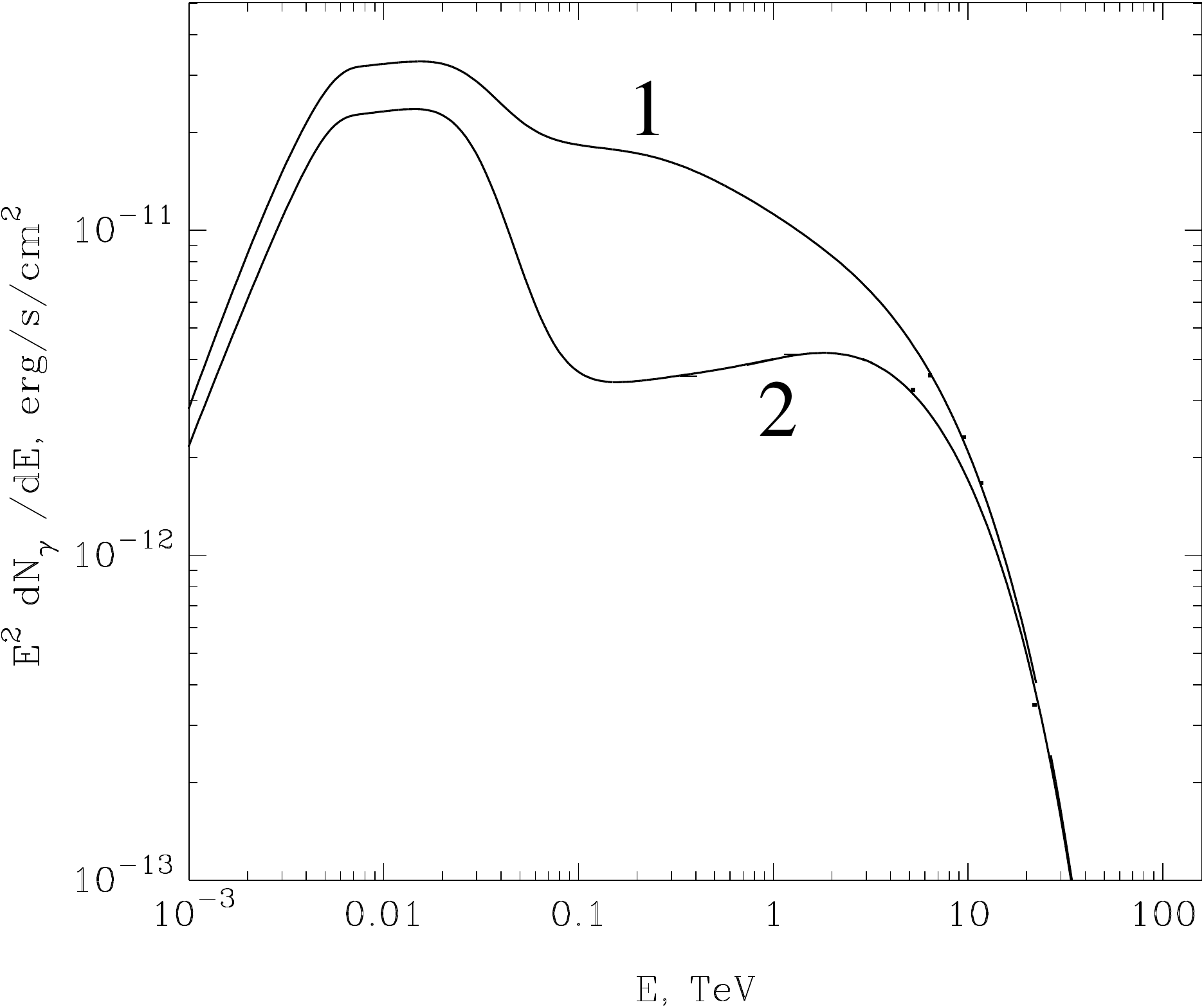}
\caption{The IC SED obtained with two different IC cross-sections. Curve 1 is calculated using the angle averaged 
inverse Compton cross-section. Curve 2 corresponds to calculations in which the angle-dependent cross-section has been 
averaged only over the orbital phase interval $0.45<\phi<0.9$, roughly when the emitter is between the companion 
star and the observer (i.e. $\theta_{\rm ic}$ relatively small; see Sec.~\ref{Sec_results}). 
The calculations were performed for $Z_0=2\cdot10^{12}$~cm, $B=0.1(Z/Z_0)^{-1}$~G, $\eta=35$,  
$i=25^\circ$, and $V_{\rm adv}=10^{10}$~cm/s. We show added both the jet and the counter-jet component.}
\label{fig:avpr}
\end{figure}

Finally, we want to emphasise the impact of the angular dependence of the IC scattering on the observed
spectrum. The importance of this effect for the IC SED is demonstrated by Fig.~\ref{fig:avpr}. The two curves shown in this figure are
obtained { under different assumptions}. Whereas curve 1 is calculated using the angle-averaged Compton cross-section, curve 2 is calculated
using the precise angular dependent cross-section and the exact interaction geometry, being averaged over the orbital phase interval 
$0.45<\phi<0.9$ (see Sec.~\ref{Sec_results}). 

%

\section{Results}
\label{Sec_results}

{ We now apply the model developed in the previous section to \source{}}. We have computed the SED and the orbital lightcurves for
different relevant situations. Some of { the SED presented here} are calculated for the inferior conjunction of the compact
object, i.e. $\phi=0.72$ (infc; when the compact object is between the observer and the companion) and the superior conjunction, i.e.
$\phi=0.06$ (supc; when the compact object is behind the companion star with respect to the observer).
\hess{} has reported on the temporal and spectral characteristics of the emission from \source{} 
\citep{hess_ls5039_06}.
Because of the lack of statistics, \hess{} has reported the SED for two different phase intervals:{  $0.45<\phi<0.9$~($\sim$infc);
$\phi<0.45$~and~$\phi>0.9$~($\sim$supc)}. In order to explore the model parameter space by comparing the calculations with the \hess{} data, some of { the
calculated SED are averaged} over the same phase intervals.

\subsection{Exploring the model parameter space}

In this section, we fix $B_0=0.05$~G, which is weak enough { to provide a radiation dominated environment} for $\sim$~TeV
electrons (see Eq.(\ref{eq:b_hard})). We also assume a high acceleration efficiency with $\eta=10$, required by the extension of
the \hess{} reported spectrum well beyond $10$~TeV (see Fig.~\ref{fig:contour}). Given the discussions in previous sections, we
assume the following values for the remaining parameters: $Z_0=10^{12},\,2\cdot10^{12},\,5\cdot10^{12}$~cm; $V_{\rm
adv}=10^9,\,10^{10}$~cm/s; and $i=15^\circ,\,25^\circ,\,55^\circ$. We compute the gamma-ray { SED for infc and supc}. These are
two distinct phases with very different physical conditions for  gamma-ray production and absorption. For instance, the maximum
gamma-ray absorption (and the minimum  $E_{\rm th}$ of this process) occurs in supc.

%
\begin{figure}
\includegraphics[width=0.5\textwidth]{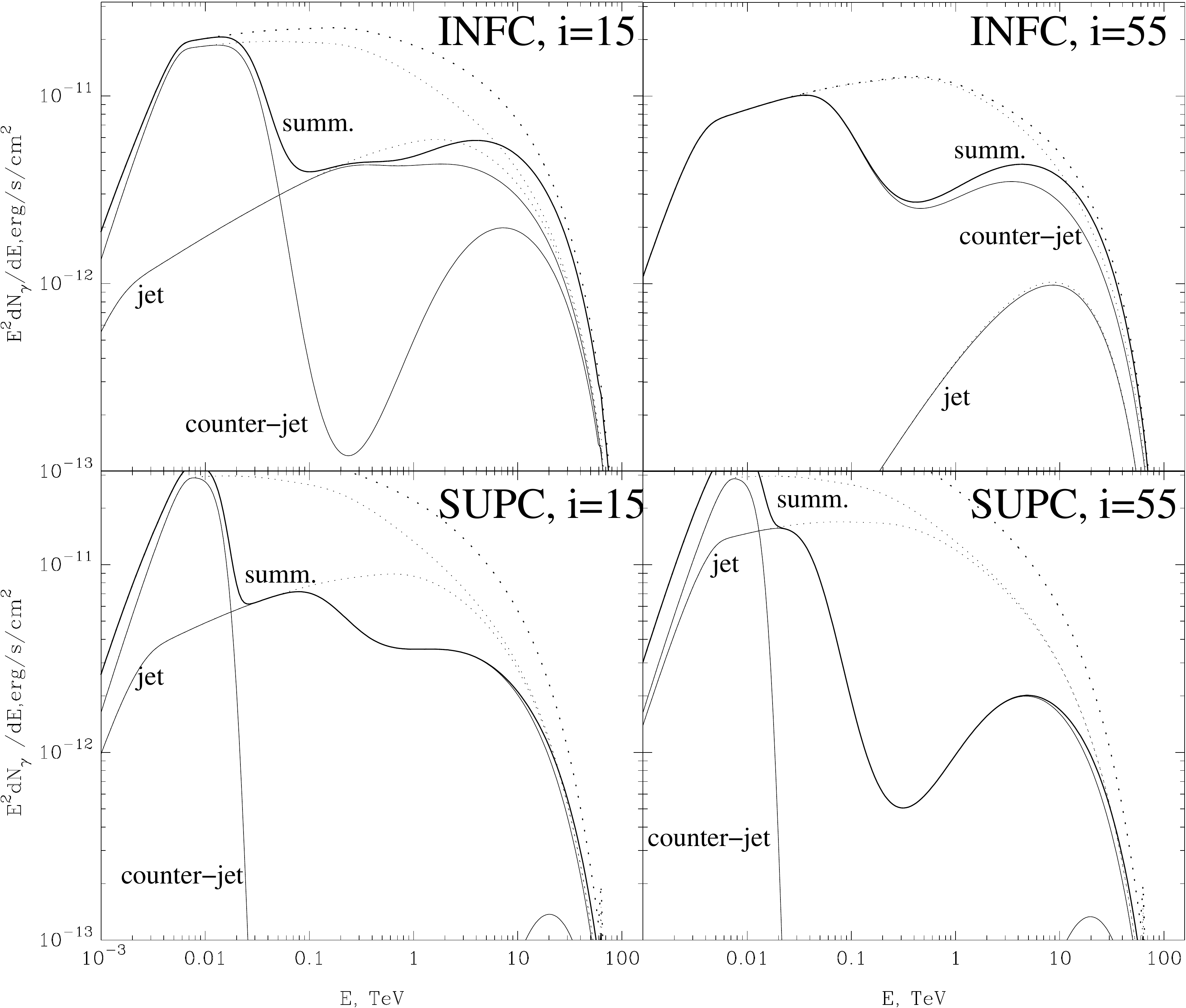}
\caption{Gamma-ray SED for supc (bottom panels) and infc (top panels) for different inclination angles: $i=15^\circ$ (left panels) 
and $55^\circ$ (right panels). 
The remaining parameter values are $Z_0=2\cdot10^{12}$~cm, $V_{\rm adv}=10^{9}$~cm/s, $\eta=10$ and 
$B=0.05(Z/Z_0)^{-1}$~G. The jet and the counter-jet components, plus the summation of both, are shown. 
The absorbed components are presented with solid lines, and the production components are shown with dotted lines.}
\label{fig:slow_inclination}
\end{figure}

\begin{figure}
\includegraphics[width=0.5\textwidth]{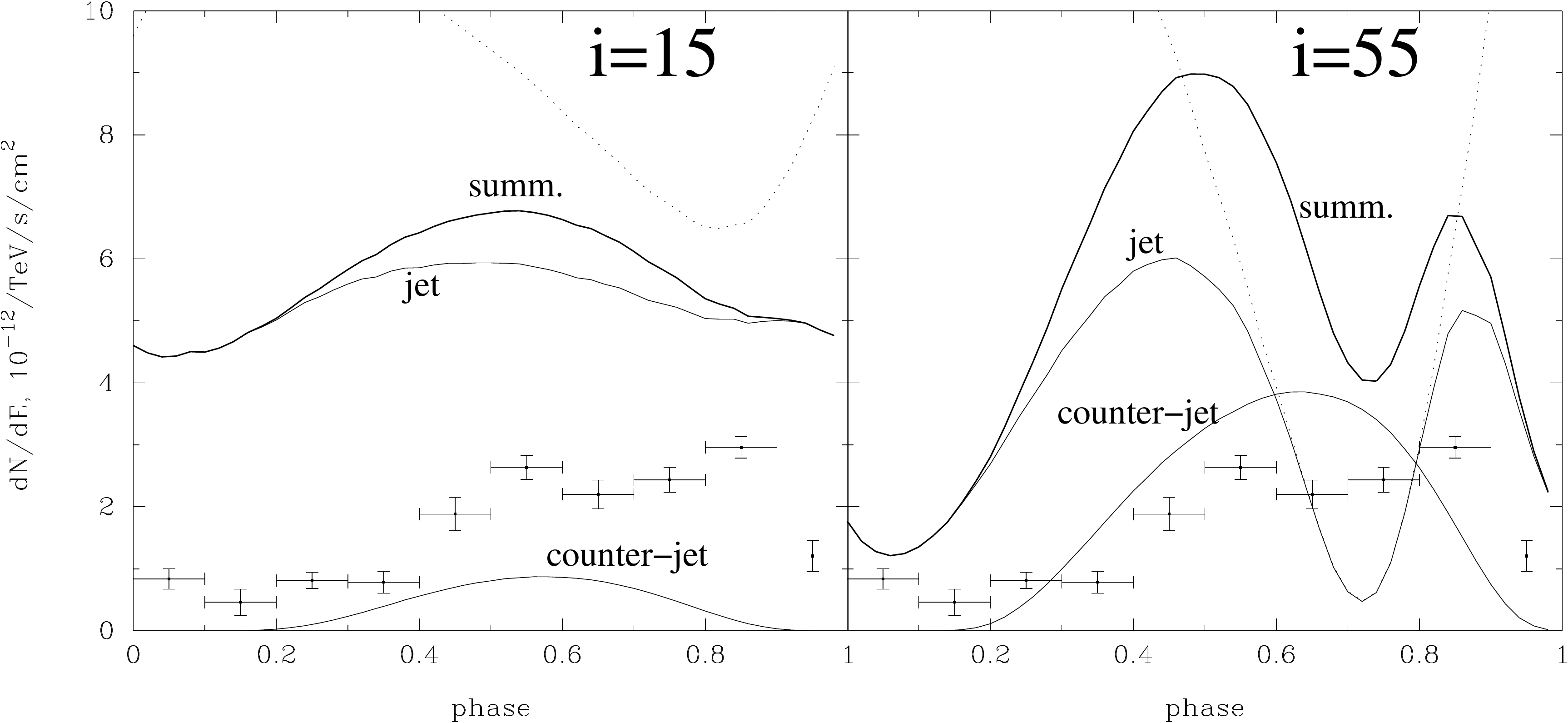}
\caption{Gamma-ray orbital lightcurves for the same cases and parameter values as in 
Fig.~\ref{fig:slow_inclination}.
The jet and the counter-jet components, plus the summation of both, are shown. 
The absorbed components are presented with solid lines, and the production components are shown with dotted lines. 
We note that the counter-jet production component 
is not seen since it is off scale.}
\label{fig:slow_inclination_lc}
\end{figure}

In Fig.~\ref{fig:slow_inclination}, we show the gamma-ray SED when the electron propagation is negligible, i.e. $V_{\rm adv}=10^9$~cm/s, for
two different $i=15^\circ,\,55^\circ$, and $Z_0=2\cdot10^{12}$~cm. This relatively slow advection velocity implies that the gamma-rays are
produced at $Z_0$. As seen from Fig.~\ref{fig:slow_inclination}, for small inclination angles (left panels), there are no strong spectral
differences between supc and infc, and the orbital lightcurve is rather flat, { as shown in Fig.~\ref{fig:slow_inclination_lc} (left
panel)}. This is caused by the small changes in the angle of
interaction for $i=15^\circ$. We note that the orbital lightcurve { would peak more strongly around apastron passage} for slightly larger 
$\eta$,
because of the strong $\eta$-dependence of the highest energy gamma-rays for the adopted $Z_0$. We note that the counter-jet impact is not
strong for low inclination angles but below $\sim 10$~GeV, because its VHE emission is strongly absorbed. 
This jet dominance is a purely geometrical effect in the IC interaction and gamma-gamma absorption. 
For high inclination angles, the interaction and absorption angles
change significantly { yielding strong spectral and flux variations} along the orbit, as shown in Fig.~\ref{fig:slow_inclination_lc} (right
panel). Two remarkable features are a double-bump orbital lightcurve, which is already significant at $i\approx 25^\circ$, and a more
relevant VHE counter-jet for $i\ga 25^\circ$ due to weaker gamma-ray absorption for this component.

\begin{figure}
\includegraphics[width=0.5\textwidth]{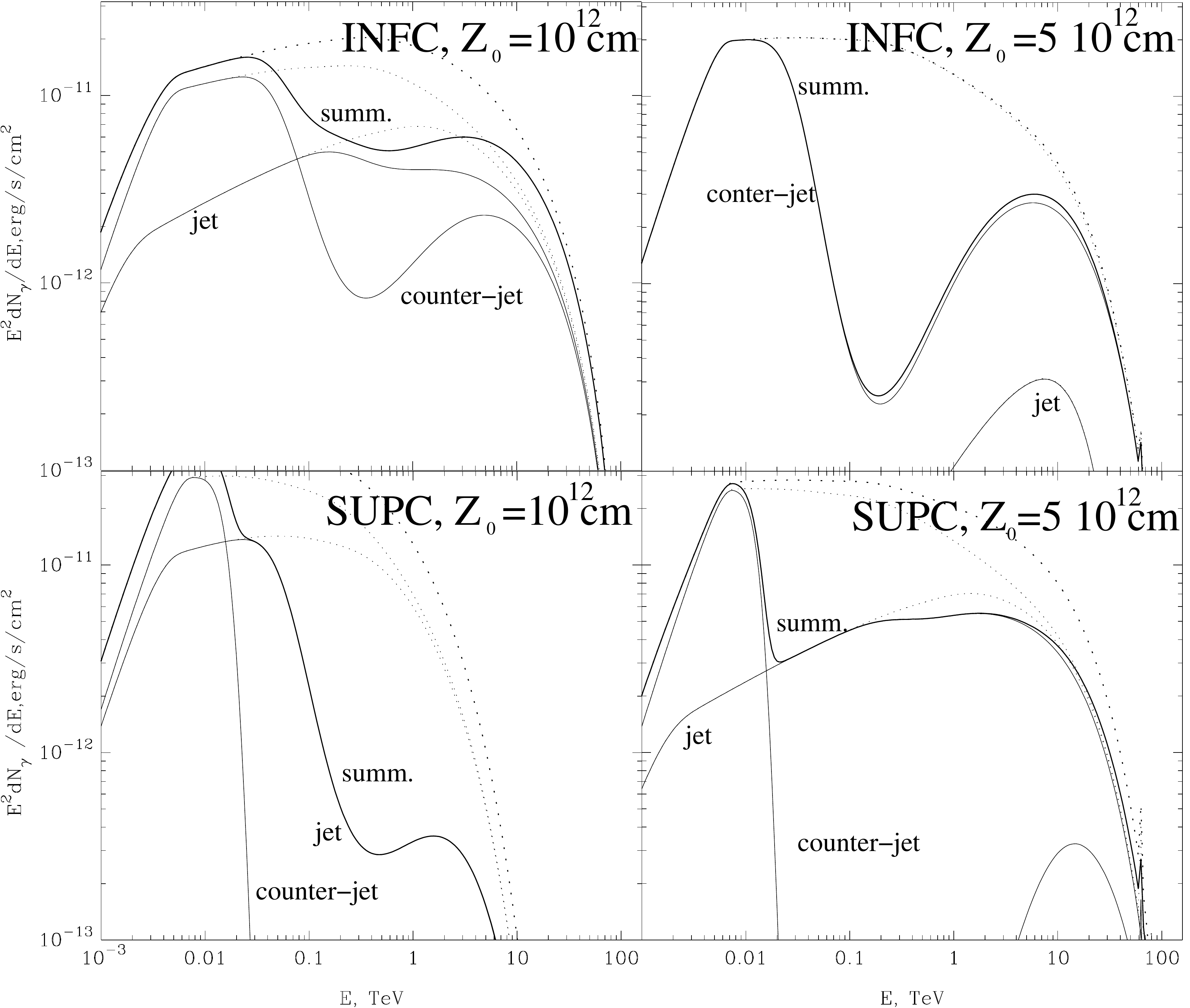}
\caption{Gamma-ray 
SED for supc (bottom panels) and infc (top panels) for different $Z_0=10^{12}$~cm (left panels) and 
$5\cdot10^{12}$ (right panels). 
The rest of parameters are assumed to be $i=25^\circ$, $V_{\rm adv}=10^{9}$~cm/s, 
$\eta=10$, and $B=0.05(Z/Z_0)^{-1}$~G. The jet and the counter-jet components, plus the summation of both, are shown. 
The absorbed components are presented with solid lines, and the production components are shown with dotted lines.}
\label{fig:slow_zeta}
\end{figure}

\begin{figure}
\includegraphics[width=0.5\textwidth]{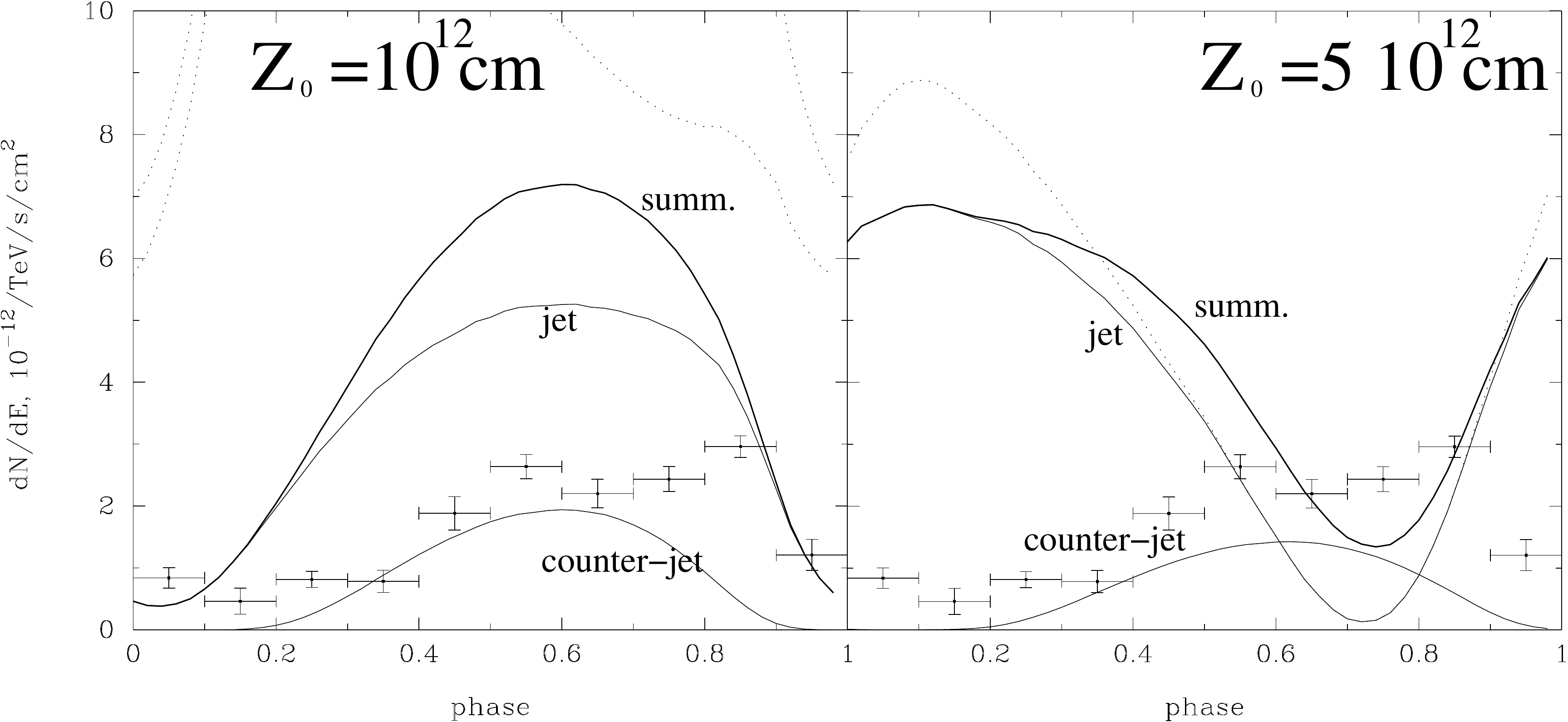}
\caption{Gamma-ray orbital lightcurves for the same cases and parameter values as in 
Fig.~\ref{fig:slow_zeta}. The jet and the counter-jet components, plus the summation of both, are shown. 
The absorbed components are presented with solid lines, and the production components are shown with dotted lines.
We note that the counter-jet production component for 
$Z_0=5\cdot 10^{12}$~cm { is not seen since it is off scale}.}
\label{fig:slow_zeta_lc}
\end{figure}

For an advection velocity $V_{\rm adv}=10^9$~cm/s and a fixed $i=25^\circ$, the dependence of the results on $Z_0$ is demonstrated by
Fig.~\ref{fig:slow_zeta}. For  $Z_0=10^{12}$~cm (left panels), fast IC energy losses around periastron prevent efficient electron
acceleration beyond 1~TeV unless $\eta<10$. A result of this effect is a strong steepening in the highest energy part of the unabsorbed
SED at supc, and as a dip in the orbital lightcurve around the same phase,{  as shown in Fig.~\ref{fig:slow_zeta_lc} (left panel)}. For $Z_0=10^{12}$~cm, 
gamma-ray absorption { deforms
the observed SED and the orbital lightcurve strongly,} producing a maximum in the lightcurve around infc, $\phi=0.72$, although 
orbital changes of $\theta_{\rm IC}$ shift the maximum of the emission to slightly earlier phases. For larger $Z_0$, e.g. 
$Z_0=5\cdot 10^{12}$~cm
(Fig.~\ref{fig:slow_zeta}, right panels), the effect of gamma-ray absorption becomes less relevant, thus only the IC angular dependence 
determines the shape of the orbital lightcurve yielding a pronounced dip around infc (see
Fig.~\ref{fig:slow_zeta_lc}, right panel). In this case, the contribution of
the counter-jet is significant around infc. 
%
\begin{figure}
\includegraphics[width=0.5\textwidth]{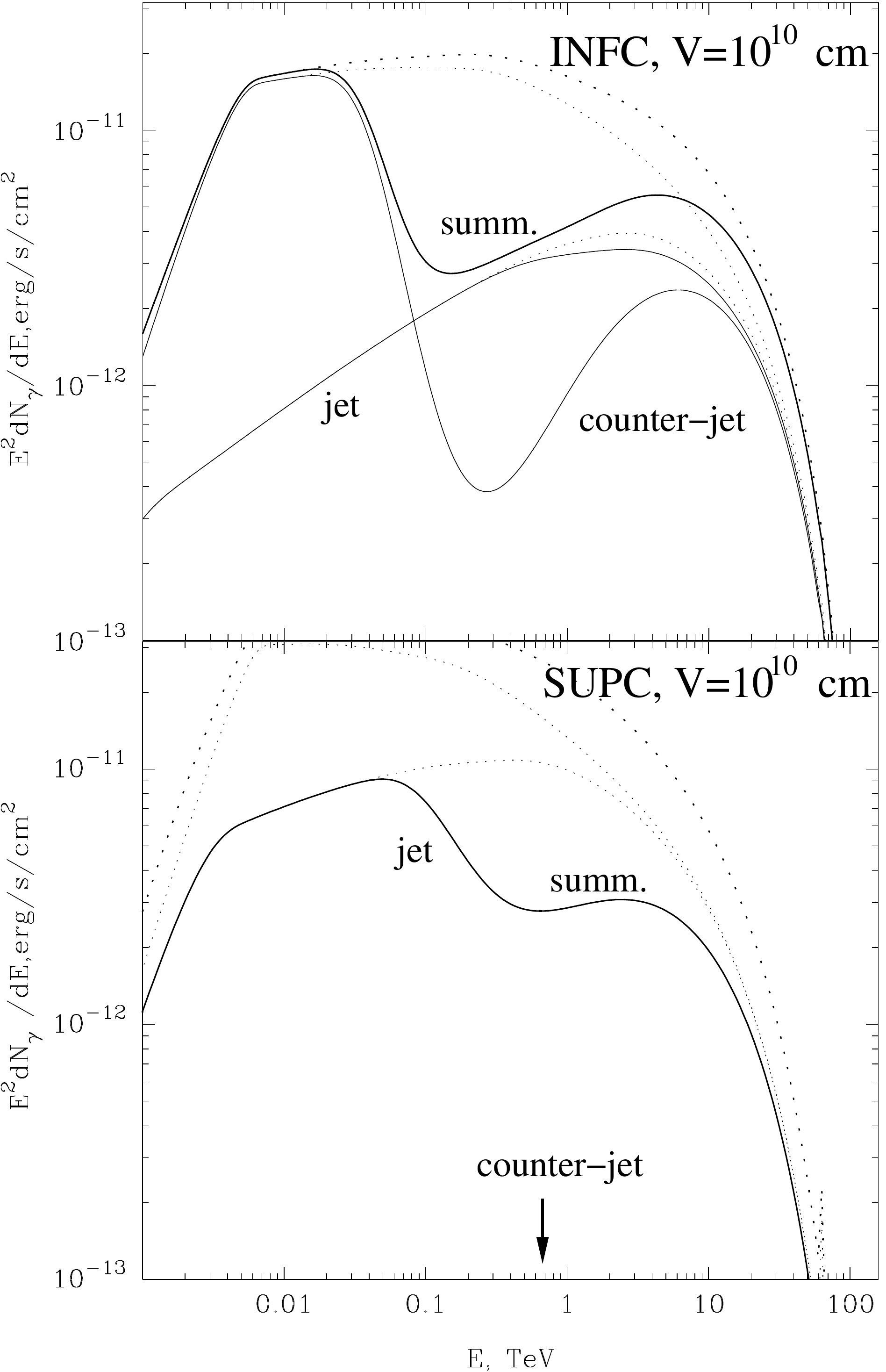}
\caption{Gamma-ray 
SED for supc (bottom panel) and infc (top panel) for $V_{\rm adv}=10^{10}$~cm/s. 
The remaining parameter values are $i=25^\circ$, $Z_0=2\cdot10^{12}$~cm, $\eta=10$, and $B=0.05(Z/Z_0)^{-1}$~G. 
The jet and the counter-jet components, plus the summation of both, are shown. 
The absorbed components are presented with solid lines, and the production components are shown with dotted lines.}
\label{fig:slow_fast}
\end{figure}

\begin{figure}
\includegraphics[width=0.5\textwidth]{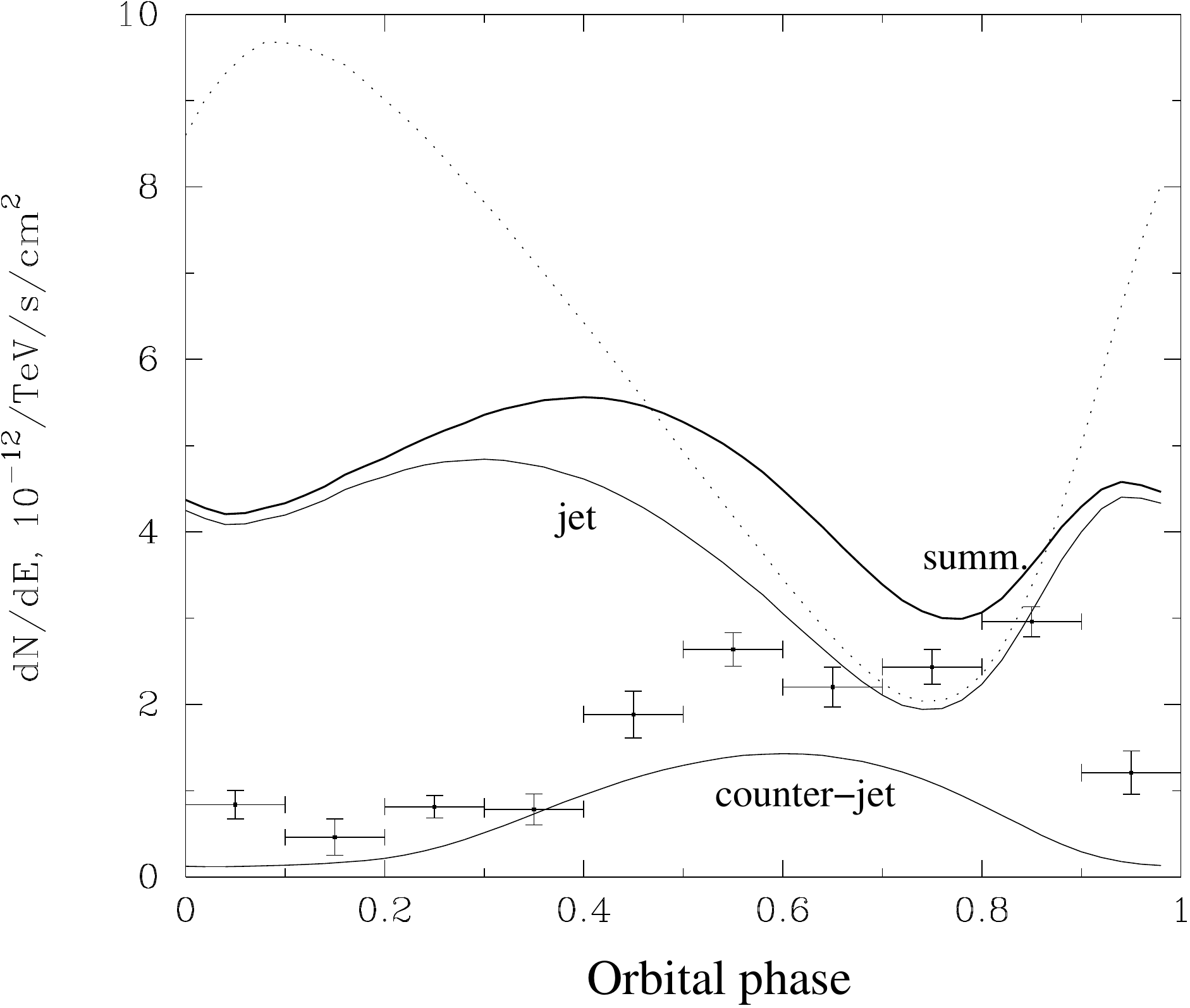}
\caption{Gamma-ray orbital lightcurve for the same case and parameter values as in 
Fig.~\ref{fig:slow_fast}. The jet and the counter-jet components, plus the summation of both, are shown. 
The absorbed components are presented with solid lines, and the production components are shown with dotted lines.}
\label{fig:slow_fast_lc}
\end{figure}

\begin{figure}
\includegraphics[width=0.5\textwidth]{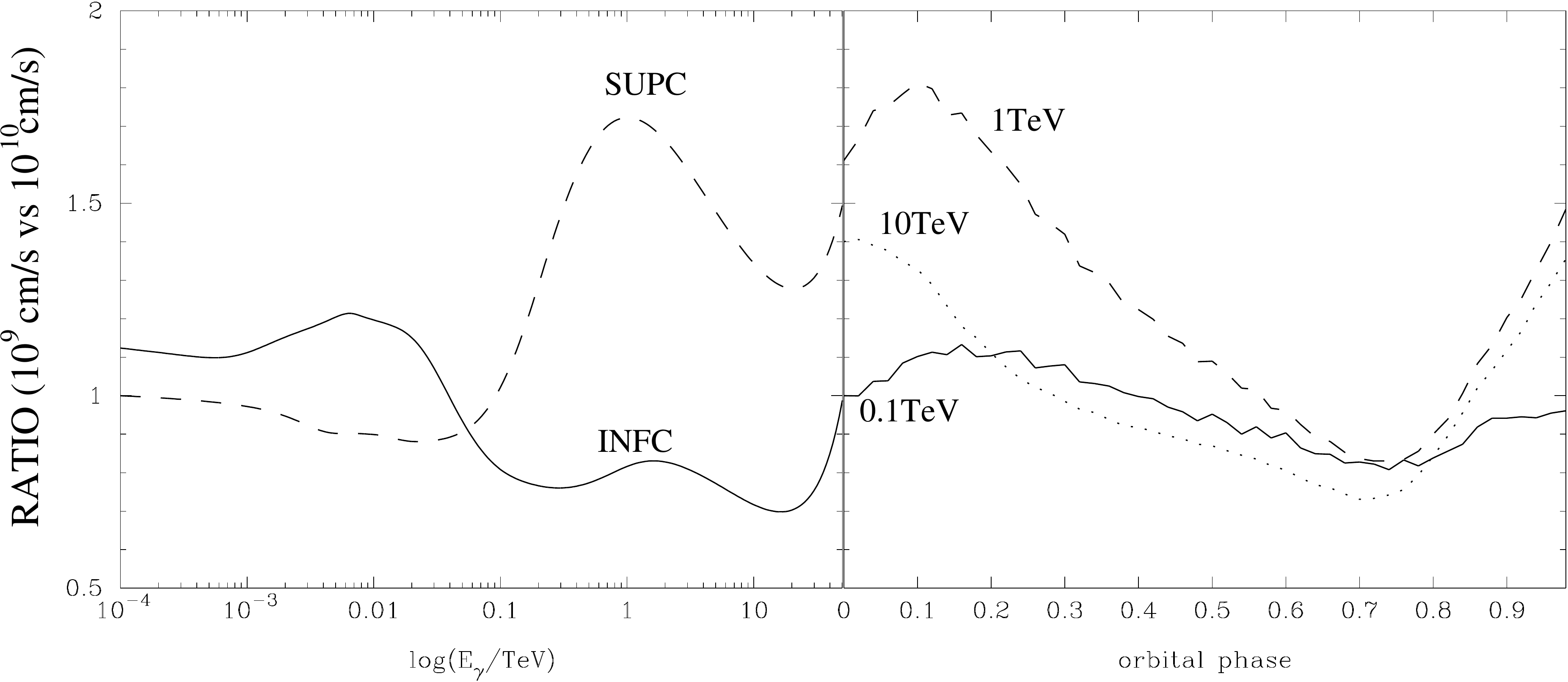}
\caption{In the left panel, the
ratio fast flow case ($V_{\rm adv}=10^{10}$~cm/s) to slow flow
case ($V_{\rm adv}=10^9$~cm/s) for the SED averaged over two orbital phase intervals:
$0.45<\phi<0.9$ (solid line -INFC-); $\phi<0.45$ and $\phi>0.9$ (dashed line -SUPC-). 
We show added both the jet and the counter-jet component.
In the right panel, the ratio for the same two $V_{\rm adv}$ values for the orbital lightcurves at three
different energies (0.1 -solid line-, 1 -dotted line- and 10~TeV -dashed line-).
The parameter values are the same as in Fig.~\ref{fig:slow_fast}.}
\label{fig:slow_fast_ratio}
\end{figure}

We compute the gamma-ray SED and orbital lightcurves for an order of magnitude faster advection velocity, i.e.
$V_{\rm adv}=10^{10}$~cm/s. The other parameters are fixed to $Z_0=2\cdot10^{12}$~cm,$B=0.05(Z/Z_0)^{-1}$~G, $\eta=10$, and $i=25^\circ$. The results are presented in
Figs.~\ref{fig:slow_fast} and \ref{fig:slow_fast_lc}. To show the impact of $V_{\rm adv}$, in Fig.~\ref{fig:slow_fast_ratio}
we show the ratios fast flow case ($V_{\rm adv}=10^{10}$~cm/s) to slow flow case ($V_{\rm adv}=10^9$~cm/s) for the SED (left panel), 
and orbital lightcurves at different
energies ($E_{\gamma}=0.1,\,1$ and 10~TeV; right panel). { The SED are averaged} 
over two orbital phase intervals: $0.45<\phi<0.9$; $\phi<0.45$ and $\phi>0.9$. For the lightcurves, we focus
on the jet radiation because the electron propagation has a weak impact on the counter-jet radiation. As discussed in Sec.~\ref{sec:propagation}, propagation allows
$1-10$~TeV electrons to reach regions for which $\tau_{\gamma\gamma}$ becomes quite small. This leads to { a hardening} of the photon spectrum around supc ($\Delta\Gamma\sim 0.2$).
Different patterns of variability are seen for different energy bands. We note that for comparison, the three curves are normalised to the same level. At higher
energies the emission becomes less variable because the angular dependence of the IC and pair-production cross-sections is weaker than at lower energies.

\subsection{Explaining the phase averaged energy spectra of \source}

Given the poor time resolution of the experimental data concerning the energy spectrum in \source{} ({ i.e. high statistic spectra} can be only
given for two wide orbital phase intervals), we do not intend to make { a fit to the \hess{} data.} Nevertheless, below we show that, with a certain
choice of key model parameters, we can explain the orbital phase averaged spectra reported by \hess. 

\begin{figure}
\includegraphics[width=0.5\textwidth]{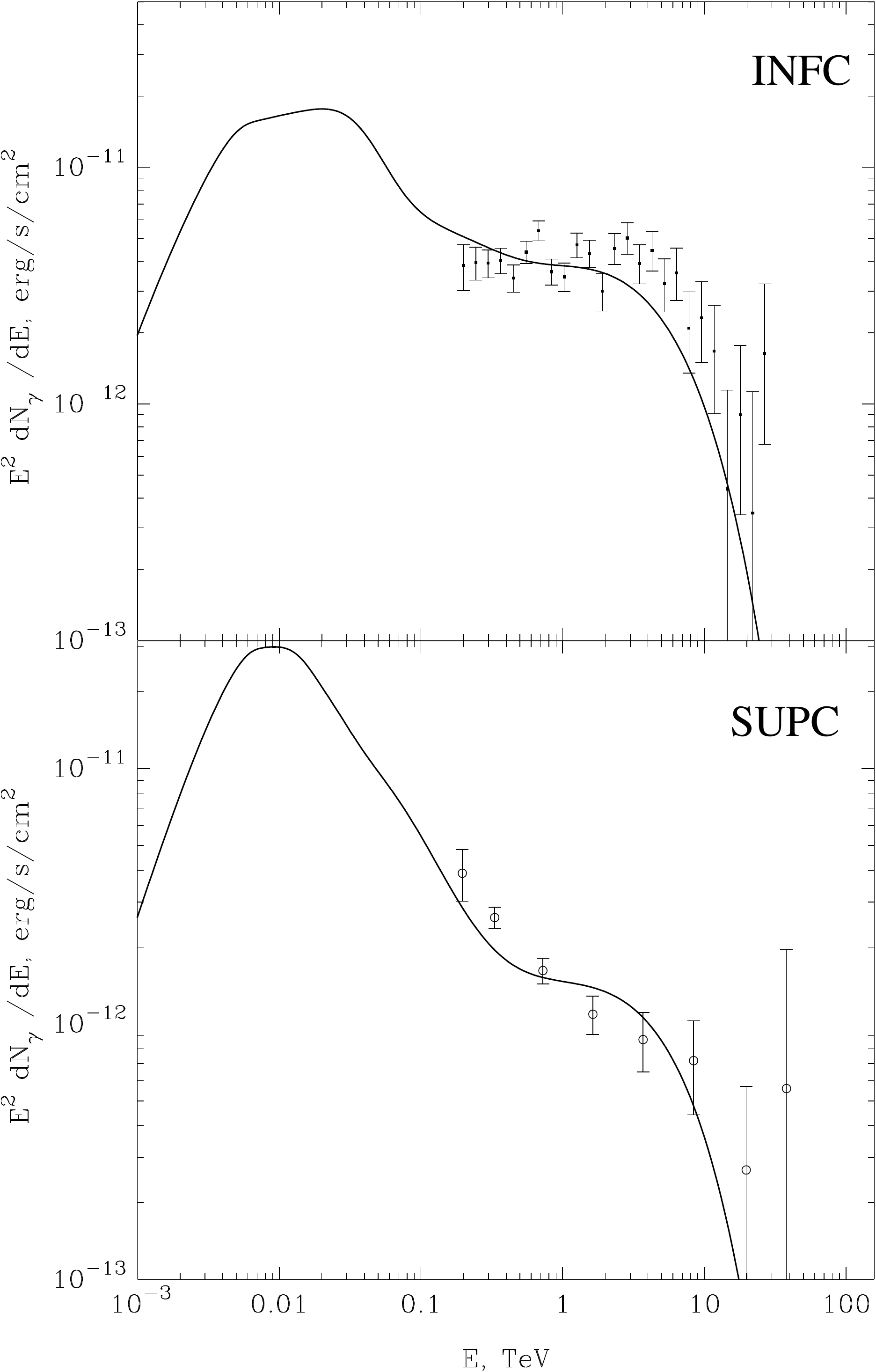}
\caption{Comparison between model results and 
\hess{} observations for both orbital phases intervals: $\sim$infc (upper panel) and $\sim$supc 
(lower panel). The following parameters are adopted: 
$i=25^\circ$; $Z_0=1.2\cdot 10^{12}$~cm; $\eta=35$; $B=0.1(Z/Z_0)^{-1}$; $V_{\rm adv}=10^{10}$~cm/s.
We show added both the jet and the counter-jet component.}
\label{fig:hess_fit}
\end{figure}

In Fig.~\ref{fig:hess_fit}, we present the \hess{} spectra for two orbital phase intervals ($0.45<\phi<0.9$; $\phi<0.45$ and $\phi>0.9$)
together with the gamma-ray SED calculated for the following parameters: $V_{\rm adv}=10^{10}$~cm/s, $\eta=35$, $B_0=0.1$~G, $i=25^\circ$,
$Z_0=1.2\cdot10^{12}$~cm, and an electron acceleration power $L_{\rm inj}=10^{35}$~erg/s. As is seen in Fig.~\ref{fig:hess_fit}, { the calculated SED agree reasonably} with the data reported by \hess{} for both phase 
intervals. Our model shows changes
in the flux by a factor $\sim 2-3$ along the orbit, similar to the observed ones, and the emission tends to peak around infc. Nevertheless, we note that
the lightcurves presented by \hess\ are derived from fits to the specific flux at 1~TeV. Given the uncertainty of the real observed spectrum,
it may well be that the \hess\ {\it fit} lightcurve would be under- or overestimated at that energy, thus we do not carry out a comparison between the observed and calculated lightcurves.
We note that some other sets of parameter values could also yield a reasonable agreement with \hess\ data. Unfortunately, because of the
orbital phase averaging, \hess\ data do not allow us to fix robustly the parameter space. This could be done with time resolved energy
spectra obtained within narrow orbital 
phase intervals ($\Delta\phi\ll 0.1$). For similar reasons, we do not show in Fig.~\ref{fig:hess_fit} the
data points of the EGRET source associated to \source{} \citep{paredes00} because the data were taken in different epochs and do not  contain
orbital phase information. Nevertheless, it is interesting to note that the observed (time averaged) ratio $\sim$GeV (EGRET) to $\sim$TeV  (\hess) 
fluxes, of about 10, is close to the model results.

%
%

\subsection{Accelerator deep inside the system}

\begin{figure}
\includegraphics[width=0.5\textwidth]{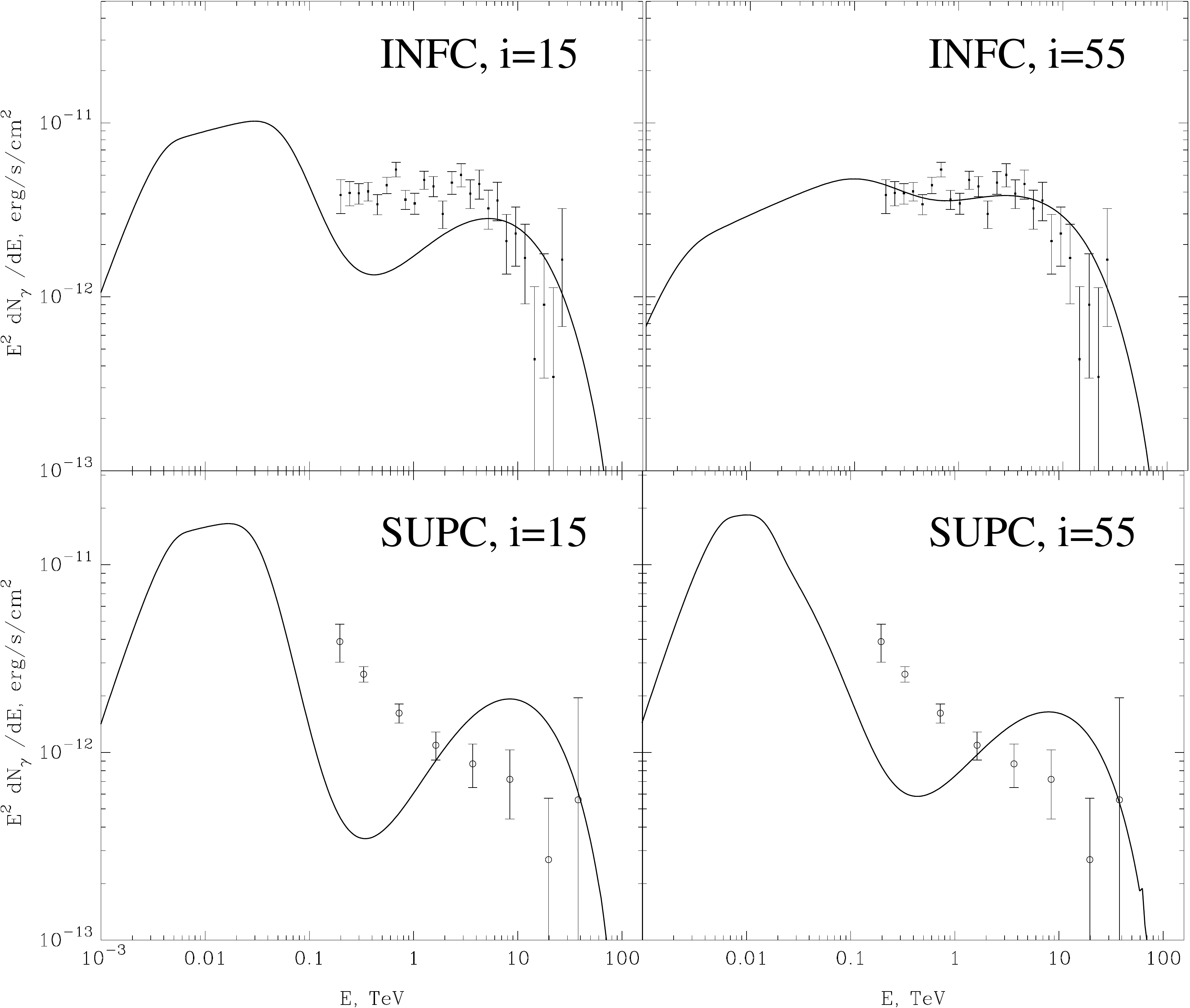}
\caption{The case of a deep accelerator. Gamma-ray SED averaged over the $\sim$supc (bottom panels) and the $\sim$infc 
(top panels) phase
intervals. Two inclination angles are adopted: $i=15^\circ$ (left panels) and $55^\circ$ (right panels). The remaining parameter values 
are $Z_0=0$, $V_{\rm adv}=10^{9}$~cm/s, $\eta=1$, and $B=0.1(Z/Z_0)^{-1}$~G. 
Here, only one component (close to the compact object) 
is computed.}
\label{fig:deep}
\end{figure}

In this section we explore the case when the accelerator is deep inside the system, i.e. $Z_0\rightarrow 0$. As discussed in
Sec.~\ref{sec:acceleration},
only an extreme accelerator with $\eta<10$ can provide a gamma-ray spectrum extending to $\ga 10$~TeV. In Fig.~\ref{fig:deep}, we show
{ the SED calculated} averaging over two orbital phase intervals ($0.45<\phi<0.9$; $\phi<0.45$ and $\phi>0.9$), taking $i=15,~55^\circ$, and
$B_0=0.1$~G. This case does not depend on the advection velocity. As seen in the figure, we note that the spectrum around supc, i.e.
$\phi<0.45$ and $\phi>0.9$, { can be barely reproduced} for any inclination angle, whereas calculations for the phases around infc,
$0.45<\phi<0.9$, are in reasonable agreement with \hess{} data, in particular for $i=55^\circ$. 

\begin{figure}
\includegraphics[width=0.5\textwidth]{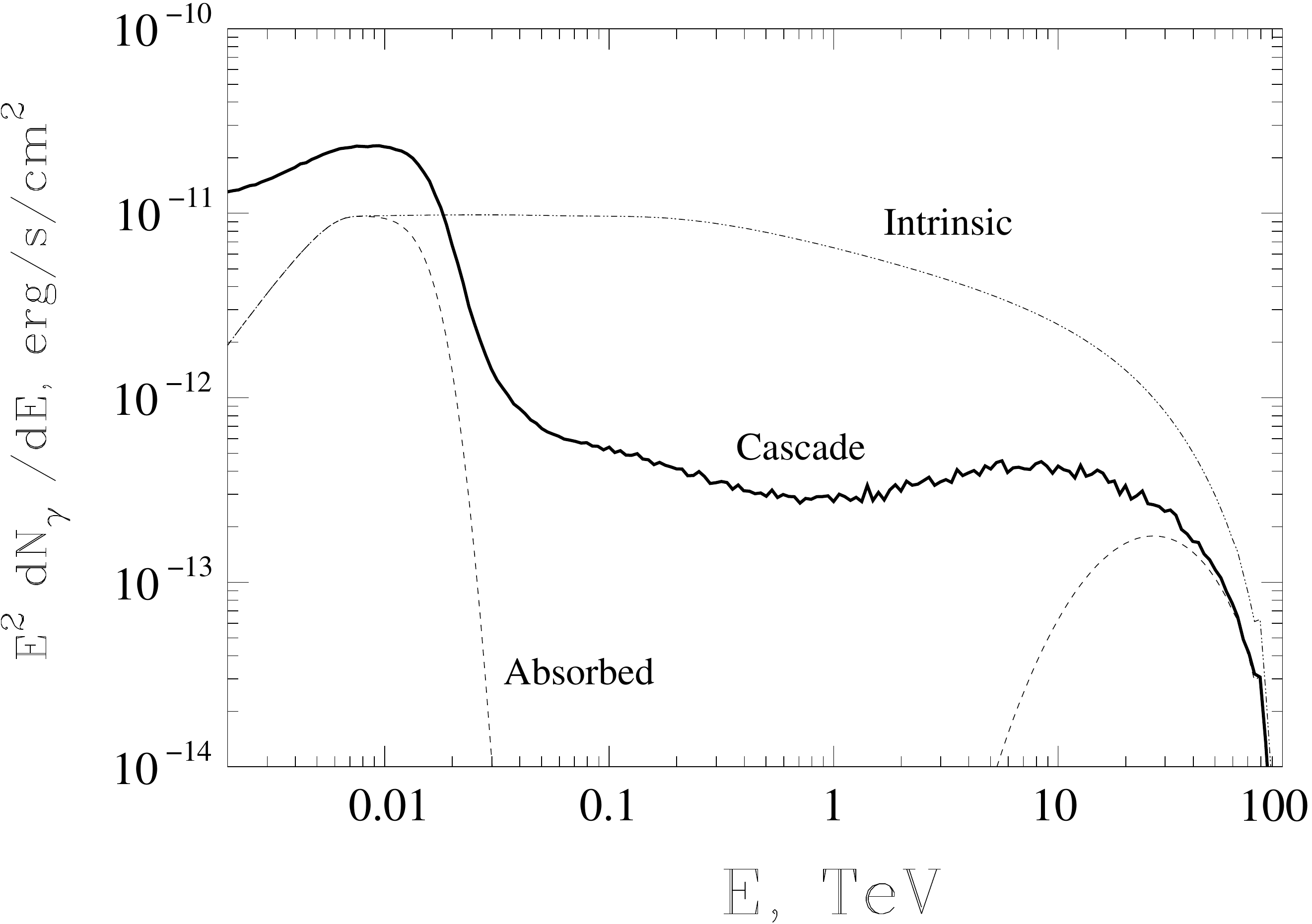}
\caption{Gamma-ray SED obtained from a Monte Carlo calculation of an electromagnetic cascade for the emission right at supc 
(non-averaged) in the deep accelerator case (see Fig.~\ref{fig:deep}). Here, only one component (close to the compact object) 
is computed.}
\label{fig:cascade}
\end{figure}

A deep accelerator implies very large optical depths ($\tau_{\gamma\gamma}\gg 1$) for photons produced all along the orbit. This leads to an
amount of absorbed energy much larger than the one observed in VHE gamma-rays. However, this energy does not disappear, but it is released at
other wavelengths. In particular, if the magnetic field in the system is small enough, $\ll 10$~G (see Sec.~\ref{sec:secondary}), absorption
of gamma-rays will initiate an electromagnetic cascade \citep{aharonian06,bednarek07} in the stellar radiation field. In case of a fully
developed cascade ($\tau \gg 1$), the  SED is characterised by a standard shape with a maximum at the threshold  $E_{\rm th} \sim m_e^2c^4/kT
\sim 10 \ \rm GeV$, a sharp drop above $E_{\rm th}$, and a flat part at $E_\gamma\ga 10 E_{\rm th}$. Under an almost monoenergetic, e.g. black-body
type, radiation field, the optical depth decreases to larger energies, thus the source becomes optically thin and the energy spectrum at the
highest energies is dominated by the unabsorbed intrinsic component. All these features can be seen in Fig.~\ref{fig:cascade}, where 
the cascade SED has been calculated for the supc (non-averaged) case. In \source,
the cascade development yields an order of magnitude or { larger GeV flux compared to the TeV energy flux}. Whereas the flat SED of the
cascade radiation could explain the HESS spectral points in the infc phase interval, { it hardly matches} the steeper spectrum for the supc
phase interval. 

\begin{figure}
\includegraphics[height=0.5\textwidth,angle=270]{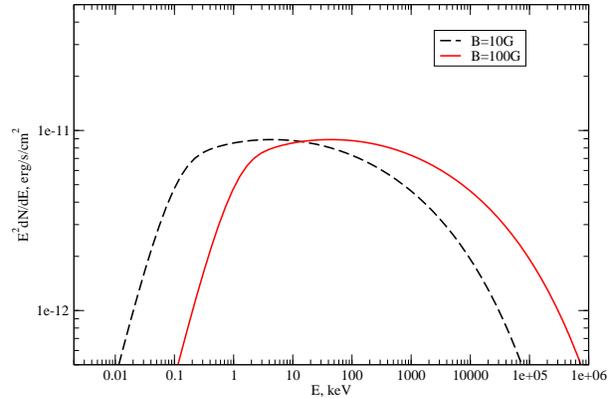}
\caption{Synchrotron radiation of secondary electrons for two magnetic fields: $B=10$ (dashed line),~$100$~G (solid line).}
\label{fig:synch}
\end{figure}

If the magnetic field in \source{} exceeds 10~G, the cascade will be effectively suppressed because of dominant synchrotron cooling of the
secondary electrons created everywhere  in the system. In such a case, { the main fraction of nonthermal energy} will be released in the
X-ray/soft gamma-ray domain. This is shown in Fig.~\ref{fig:synch}, in which the synchrotron radiation is shown for two values of the
system average magnetic field: 10 and 100~G. We note that the fluxes shown in Fig.~\ref{fig:synch} are calculated for an acceleration power
of primary electrons $L_{\rm inj}=10^{35} \ \rm erg/s$, required to explain the TeV fluxes reported by \hess{} (see Fig.~\ref{fig:deep}).
For a system average magnetic field of $\sim 10$~G, the synchrotron radiation of secondary electrons peaks around 10~keV, with a flux
which is close to the observed fluxes  \citep{bosch07}. In the case of a stronger magnetic field (e.g. 100~G), the synchrotron
emission { peaks in the soft gamma-ray range,} being also  below the reported fluxes \citep{strong01}.

%
%

%

%

\section{Summary}

In this paper we present the results of a detailed numerical study of the spectral and temporal properties of the high energy gamma-ray
emission from the binary system  \source{} adopting an IC scenario. The calculations have been conducted in the framework of a specific
model which assumes that gamma-rays are produced in a jet-like structure perpendicular to the orbital plane. However, the obtained results
and conclusions concerning the electron accelerator and the gamma-ray emitter have broader implications. { The sheer fact} that we see a
modulated  gamma-ray signal with { a $3.9$~day period}, with a hard energy spectrum extending to 10~TeV and beyond, implies that: (a) the
gamma-ray emitter is located close to the binary system, i.e. at distances comparable to $R_{\rm orb}$; and (b) we deal with an efficient
accelerator with { acceleration rate not far from} the theoretically maximum possible rate, i.e. $\eta<100$ 
($\eta<10$ for an accelerator located deep inside the binary system). 
Another interesting model-independent conclusion is { that if the  gamma rays have an inverse Compton origin},
then we can restrict the magnetic field in the gamma-ray emitter within a quite narrow range. { The fact that the IC scattering which produces the TeV photons occurs deep in the KN regime implies that the observed spectrum with photon index $\Gamma \sim 2$ requires a very hard electron  energy distribution with power-law index $\sim 1$.} Such a hard spectrum cannot be formed even { in the case of a monoenergetic} (e.g. Maxwellian type)
injection electron spectrum unless they cool through the IC scattering in the KN regime\footnote{Synchrotron/Thomson IC cooling would yield
a $\propto E_{\rm e}^{-2}$ electron energy distribution.} (see Eq.(\ref{eq:b_hard})). On the other hand the magnetic field cannot be much
smaller than 0.1 G in order to provide an adequate acceleration rate. Thus,{ the accelerator 
emitter magnetic field} should be confined in a band around 0.1 G.

Because of the close location of the gamma-ray emitter to the companion star, the modification of the spectrum of gamma-rays { due to
the gamma-gamma absorption process} is unavoidable. Moreover, gamma-ray absorption strongly depends on the orbital phase. Because of the slightly
{ elliptical orbit}, the phase-dependence of the photon density is moderate. However, the orbital change of the interaction angle { results in a change in the threshold of the interaction}, and thus leads to strong modulation of the optical depth. Interestingly, the energy-dependent
modification of the gamma-ray energy spectrum produced by gamma-gamma absorption cannot explain the data reported by HESS for two phase
intervals. In particular, whereas one should expect the strongest variation of the gamma-ray flux at $\sim 100$~GeV; 
\hess{} results show
just the opposite. Nevertheless, the anisotropy of the radiation field has a strong impact not only on the gamma-gamma absorption, but also
on the spectral characteristics of the IC scattering. We show here that the combination of these two effects, both related to the anisotropic
character of the interaction of the stellar radiation field with relativistic electrons (IC) and gamma-rays (pair production), can
satisfactorily explain the \hess{} observations reported for the supc and infc phase intervals. In addition, we note that the
phase-dependence of the electron maximum energy and the electron advection along the jet could play an important role in the formation of { the gamma-ray spectra.} 

The absorption of gamma-rays does not imply that this energy disappears. The secondary electrons interacting with the stellar radiation and
ambient magnetic fields produce secondary radiation. { There are two well defined possibilities; either the energy density of the ambient
medium is dominated by the stellar radiation field or by the magnetic field.} In the former case, pair production and further IC scattering 
initiate an electromagnetic cascade. In the latter case, synchrotron radiation is produced. As noted above, a typical ambient magnetic field
determining the boundary between these two regimes { is several Gauss. When an efficient cascade occurs}, the source becomes more transparent
at very high energies, { although the energy is mainly released around the threshold energy $\sim 10$ GeV.  In the case of the suppression} of the
cascade through synchrotron cooling, the source is less transparent, and synchrotron radiation is released mostly at keV-MeV energies, with
luminosities exceeding the VHE gamma-ray luminosity. Interestingly, a $\sim 10$~G magnetic field is plausible for the surroundings of a massive star, so 
\textit{a priori} one cannot exclude any of both possible scenarios, which in fact may be realised simultaneously in different parts of the
system. In such a case, both cascading and synchrotron radiation of secondaries { would contribute comparably} to the emerging high energy
radiation. In this regard, future deep X-ray and gamma-ray observations, together with TeV observations with more sensitive instruments 
{ capable of spectral measurements} on short time scales, will provide { insights} into the origin of the nonthermal emission of this enigmatic
source.  

\section*{Acknowledgments}
{ Authors are grateful to M.~Chernyakova, A.~Taylor and the anonymous referee for useful comments.}
V.B-R. gratefully acknowledges support from the Alexander von Humboldt Foundation. 
V.B-R. thanks MPIfK for its kind hospitality.
V.B-R. acknowledges support by DGI of MEC under grant
AYA2004-07171-C02-01, as well as partial support by the European Regional Development Fund (ERDF/FEDER). 

\hyphenation{Post-Script Sprin-ger}

\label{lastpage}
\end{document}